\documentclass[reprint,prb,aps,groupedaddress]{revtex4-1}

\usepackage{graphicx}
\usepackage{amsmath}
\usepackage{amsfonts}
\usepackage{bbm}
\usepackage{physics}
\usepackage[normalem]{ulem}

\usepackage{xcolor}

\long\def\comment#1{}

\begin{document}
\title{Dynamical freezing and switching in periodically driven bilayer graphene}
  
 \author{Soumya Sasidharan}
 \email{soumyatkmm@gmail.com}
 \author{Naveen Surendran}
\email[]{naveen.surendran@iist.ac.in}

\affiliation{Indian Institute of Space Science and Technology,
  Valiamala, Thiruvananthapuram-695547, Kerala, India}
  
\date{\today}

\begin{abstract}

A class of integrable models, such as the one-dimensional transverse-field Ising model, respond nonmonotonically to a periodic drive with respect to the driving parameters and freezes almost absolutely for certain combinations of the latter. In this paper, we go beyond the two-band structure of the Ising-like models studied previously and ask whether such unusual nonmonotonic response and near-absolute freezing occur in integrable systems with a higher number of bands. To this end, we consider a tight-binding model for bilayer graphene subjected to an interlayer potential difference. We find that when the potential is driven periodically, the system responds nonmonotonically to variations in the driving amplitude $V_0$ and frequency $\omega$ and shows near absolute freezing for certain values of $V_0/\omega$. However, the freezing occurs only in the presence of a constant bias in the driving, i.e., when $V= V'+V_0 \cos{\omega t}$. When $V'=0$, the freezing is switched off for all values of $V_0/\omega$. We support our numerical results with analytical calculations based on a rotating wave approximation. We also give a proposal to realize the driven bilayer system via ultracold atoms in an optical lattice, where the driving can be implemented by shaking the lattice. 

\end{abstract}


\maketitle
\include{reference.bib}

\section{\label{s-intro}Introduction}

Periodically driven quantum many-particle systems showcase a variety of phenomena such as nonequilibrium quantum phase transitions \cite{BasEma(12a), BasEma(12b), BasPer(14), EngBas(13)} and Floquet engineered topological phases \cite{BukDal(15), GolDal(14)}. In this paper, we focus on another remarkable aspect of coherent periodic driving known as dynamical many-body freezing (DMF) wherein a system responds nonmonotonically to variations in the driving parameters and freezes almost completely at certain combinations of the latter \cite{Das(10), BhaDas(12),RusSil(12)}. DMF is a many-body manifestation of the single-particle phenomena such as the dynamical localization of a particle moving on a lattice in the presence of an alternating electric field \cite{DunKen(86),EckHol(09)}, or the  coherent destruction of tunneling of a particle moving in a periodically driven double-well potential \cite{GroDit(91),GroHan(92)}.  

Multiple aspects of DMF have been explored in recent years. These include: effect of disorder \cite{RoyDas(15)}, the emergence of slow solitary oscillations \cite{BhaDas(12)}, effect of interactions \cite{HalSen(21)}, and switching of the response by tuning parameters in the Hamiltonian \cite{DasMoe(12)}. DMF has been experimentally demonstrated in a periodically driven Ising chain \cite{HegKat(14)}. For recent reviews on DMF, see Refs.  \onlinecite{HalDas(17),HalDas(21)}.

The question of whether quantum integrable systems could freeze under periodic driving due to coherent cancellation of transition amplitudes was first  addressed in Ref. \onlinecite{Das(10)}, using the one-dimensional 
  transverse-field Ising model (TFIM) as a concrete example. When the magnetic field is driven harmonically at high frequencies, for a fixed driving amplitude, the magnetization (which is a measure of the degree of freezing) shows a nonmonotonic dependence on the frequency. Remarkably, for certain combinations of the amplitude and frequency of the drive, the entire system freezes almost absolutely into the initial maximally polarized state.
  
  The One-dimensional Ising model is integrable and is exactly solvable via Jordan-Wigner transformation, which maps the system to a two-band free fermion system. In this paper, we look for DMF in a four-band system that has been studied extensively both theoretically \cite{AbeApa(10), KonGmi(12), OosHee(08), GosTan(13), ZhaLin(11)} and experimentally \cite{BosMcc(09), ZhaBas(08), FelMar(09)}, viz., the bilayer graphene.

  In bilayer graphene, an external electric field applied perpendicular to its plane induces a potential difference between the two layers \cite{MinSah(07),MccEdw(06),ChaHua(06)}. This results in an effective layer-dependent chemical potential term in the Hamiltonian, which opens a band gap. It has been experimentally shown that the applied electric field can be used to tune the energy gap \cite{ZhaYua(09),CasNov(07)}.
  
  In our model, we harmonically drive the layer-dependent potential ($-V$ for layer-1 and $V$ for layer-2). We find that DMF occurs only when a constant bias is added to the driving: $V(t) = V' + V_0 \cos \omega t$. For nonzero $V'$, the system responds nonmonotonically to variations in $V_0$ and $\omega$ and freezes for certain combinations of the latter. When $V'=0$, no freezing occurs for any combination of $V_0$ and $\omega$, which is in contrast to the two-band models such as the Ising model studied earlier, where freezing occurs without any bias in the driving. In the model for bilayer graphene we have studied, the bias acts as a switch for freezing.

  Dynamical freezing, in general, requires high driving frequencies and amplitudes, which may be difficult to achieve with the gate voltage in bilayer graphene. However, it has been shown that dynamical freezing can occur even at low frequencies provided the driving amplitude is above the thermalization threshold \cite{HalMoe(18)}. Another way to overcome the difficulty of achieving large values of drive parameters is to simulate the dynamics in a moving frame in which none of the couplings in the Hamiltonian is large \cite{HalSen(21)}. Here put forward a proposal to realize the bilayer system in an optical lattice where the driving can be implemented via lattice shaking. Dynamical localization in a one-dimensional optical lattice via lattice shaking has already been demonstrated experimentally \cite{LigSia(07)}.

  The rest of the paper is organized as follows. In Sec. \ref{s-bilayer}, we briefly review bilayer graphene and its energy spectrum. In Sec. \ref{s-driving}, we study the response of the system when the interlayer potential is driven periodically. In Sec.\ref{s-oplat}, we give a proposal to realize the driven bilayer system in an optical lattice via lattice shaking.  We conclude by discussing our results in Sec. \ref{s-sum}. 
  
  \section{\label{s-bilayer} Bilayer graphene}

\begin{figure}
    \includegraphics[scale=1]{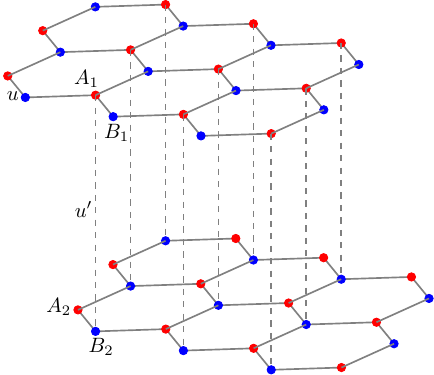}
\caption{$A$-$B$ Bilayer graphene}
\label{f-bg}
\end{figure}

Graphene is a single layer of graphite composed of carbon atoms forming a honeycomb lattice \cite{GeiNov(07), NovGei(04)}. Bilayer graphene---two connected layers of graphene---exists in two different forms\cite{WanGuo(12), RozSbo(16)}: (1) AA, in which all atoms in the top layer are placed directly above the atoms in the bottom layer, and (2) AB, in which half the atoms in the top layer, all belonging to the same sublattice, are placed above the centers of the hexagons in the bottom layer, while the atoms in the top layer belonging to the other sublattice are placed directly above the atoms belonging to one of the sublattices in the bottom layer (see Fig. \ref{f-bg}). There also exist stable structures in which one layer is rotated with respect to the other, known as twisted bilayer graphene \cite{AndMac(20),ChoKem(19), McCAbe(07)}. AB structure is more stable compared to AA and has been widely studied experimentally \cite{MooKosh(12),YanPen(11),  OulBou(17), LaiHo(08), BitVic(17)}. We consider the $ AB$ stacking first.

The unit cell contains four carbon atoms, labeled $A_1, A_2, B_1$, and $B_2$, as shown in the Fig. \ref{f-bg} 
The tight-binding model we study has two types of hopping: in-plane nearest neighbor hopping with amplitude $-u$ and inter-plane nearest neighbor hopping with amplitude $u'$ \cite{RozSbo(16)}. In addition, we also consider an external electric field applied perpendicular to the plane of the bilayer. This induces (after taking the screening effect into account) an effective voltage difference $V$ between the layers. Then the Hamiltonian is given by,\cite{RozSbo(16)}
\begin{align}
\mathcal{H} =& -u \sum_{\langle i,j \rangle}\left[a_{i,1}^\dagger  b_{j,1}+ a_{i,2}^\dagger  b_{j,2} + h.c.\right] \nonumber \\
&+ u' \sum_i \left[ b_{i,1}^\dagger  a_{i,2}+h.c.\right] \nonumber \\
&-\frac{V}{2} \sum_i \left[ a_{i,1}^\dagger a_{i,1} + b_{i,1}^\dagger b_{i,1} ] -  a_{i,2}^\dagger a_{i,2} - b_{i,2}^\dagger b_{i,2} \right],    
\label{e-bilham}
\end{align}
where $a_{i,\alpha}^{\dagger}(b_{i,\alpha}^{\dagger})$is the electron creation operator at site $i$ in sublattice A (B) belonging to the layer $\alpha$. (Here we have suppressed the spin index for notational simplicity.) In the momentum basis, the Hamiltonian becomes
\begin{align}
    \mathcal{H}_\textbf{k} &=\sum_{\textbf{k}} \Psi_{\textbf{k}}^{\dagger}H_{\textbf{k}}\Psi_{\textbf{k}},
\end{align}
where $\Psi^\dagger_{\textbf{k}}=\left[ a^\dagger_{\textbf{k}1}, b^\dagger_{\textbf{k}1}, a^\dagger_{\textbf{k}2}, b^\dagger_{\textbf{k}2}\right]$, and
\begin{align}
  H_{\textbf{k}}= 
  \begin{bmatrix}
-\dfrac{V}{2} & -u f_{\bf k} & 0 & 0 \\
-u f^*_{\bf k} & -\dfrac{V}{2} & u' & 0\\
0 & u' & \dfrac{V}{2} & -u f_{\bf k}   \\
0 & 0 &-u f^*_{\bf k}  & \dfrac{V}{2} \\ 
\end{bmatrix},
\label{e-bgham}
 \end{align}
 with
 \begin{align}
     f_{\bf k}=\exp(-ia_{0}k_{x})\left[1+2\exp(\dfrac{3ia_{0}k_{x}}{2})\cos(\dfrac{\sqrt{3}a_{0}k_{y}}{2})\right].
     \label{e-bilmomh}
     \end{align}
  Here $a_{0}$ is the distance between nearest-neighbor carbon atoms within a layer. The energy dispersions have the form:
\begin{align}
     \epsilon_{{\bf k}}^{\alpha}=\pm \dfrac{1}{\sqrt{2}}\Bigg[&\dfrac{V^{2}}{2}+u'^{2}+2u^{2}\vert f_{\bf k}\vert^{2}\nonumber \\
    & \pm u'\sqrt{-2V^{2}+u'^{2}+4u^{2}\vert f_{\bf k}\vert^{2}} \Bigg]^{\frac{1}{2}},
 \end{align}
  where $\alpha=1,2,3,4$, respectively, correspond to the choices $(+,+),(+,-), (-,+), (-,-)$. When $V=0$, the spectrum is gapless at the six corners of the hexagonal Brillouin zone, of which only two are inequivalent, which we can choose to be 
  \begin{align}
  \begin{split}
  {\bf K}&=\dfrac{2\pi}{3a_{0}}\left(1,\dfrac{1}{\sqrt{3}}\right),\\
 {\bf K^{\prime}}&=\dfrac{2\pi}{3a_{0}}\left(1,\dfrac{-1}{\sqrt{3}}\right).
 \end{split}
 \label{e-diracpts}
 \end{align}
 The chemical potential term opens up a gap in the dispersion, since for nonzero $V$, $\epsilon_{\bf k}^\alpha$ is no longer zero for any value of ${\bf k}$.

We now periodically drive the potential $V$ and study the response of the system to variations in the driving parameters.
 
\section{\label{s-driving} Periodic driving}

We choose $u'=0.2$ and $u=1$, in accordance with their experimentally determined values, \cite{RozSbo(16)} and drive the potential harmonically:
\begin{align}
    V(t) &= V_{0}\cos{\omega t}.
    \label{e-drive}
\end{align}
Our focus will be on the large amplitude and high-frequency regime, i.e., $V_0, \omega \gg u,u'$ [throughout this paper we work with units in which $\hbar = 1$]. 

To study the response to the driving, we compute the long-time average of the probability for each mode to remain in the initial state. It is useful to define the following quantities:
\begin{align}
   q_{\bf k}( t) &= \vert \bra{\psi_{{\bf k}}(0)} \ket{\psi_{{\bf k}}(t)}\vert^{2}, \label{e-qkt}\\ 
   \tilde{q}(t) &= \dfrac{1}{N}\sum_{\bf k} q_{\bf k}( t), \label{e-qtilde}\\
   \bar q_{\bf k} &= \frac{1}{T} \int_0^T q_{\bf k}( t)~ dt, \label{e-qbar}\\
    Q &=  \frac{1}{T} \int_0^T dt~ \tilde{q}(t), \label{e-qavg}
    \end{align}
where the initial state $\ket{\psi_{\bf k}(0)}$ is chosen to be the ground state corresponding to momentum ${\bf k}$ (with either one or two particles), and $N$ is the number of unit cells. Here $q_{\bf k}( t)$ is the probability for the one/two-particle state of the ${\bf k}$-mode(s) to remain in the initial state at $t$, $\tilde{q}(t)$ is the above probability averaged over all ${\bf k}$, $\bar q_{\bf k}$ is the time-average of $q_{\bf k}( t)$, and $Q$ is the latter averaged over both $t$ and ${\bf k}$.

The parameter $Q$ measures the degree of freezing, with $Q=1$ indicating absolute freezing. We first calculate $ Q$ by numerically solving the Schr\"{o}dinger equation and then understand our results within the framework of a rotating wave approximation valid at high frequencies.

We consider the system at half-filling, then, at $t=0$, the two negative energy lower bands are filled. Due to the lattice symmetry, the dynamics for a given ${\bf k}$ will be restricted to the corresponding six-dimensional two-particle sector. However, since the Hamiltonian is noninteracting, the two-particle dynamics will be determined by the dynamics of the one-particle sector, which we consider first.

\subsection{\label{s-1p} One-particle sector}

At $t=0$, we occupy the lowest energy state for each ${\bf k}$ with one particle. Since $V_0\gg u,u'$, we can approximate the initial state to be the ground state of $H_{\bf k}$ [Eq. (\ref{e-bgham})] with $u=u'=0$. Then, 
 \begin{align}
   \ket{\psi_{\bf k}(0)}=\dfrac{1}{\sqrt{2}}
   \begin{bmatrix}
     1\\
    \dfrac{\vert f_\textbf{k}\vert}{f_{\bf k}}\\
    0\\
   0\\
    \end{bmatrix}.
   \label{e-Gsbg}
  \end{align}     
Starting with the above state, we let the system evolve under the periodic drive [Eq. (\ref{e-drive})] and numerically evaluate $q_{\bf k}(t)$ for varying driving frequency $\omega$, keeping the amplitude $V_0$ fixed. 

\begin{figure*}
\includegraphics[width=0.8\textwidth]{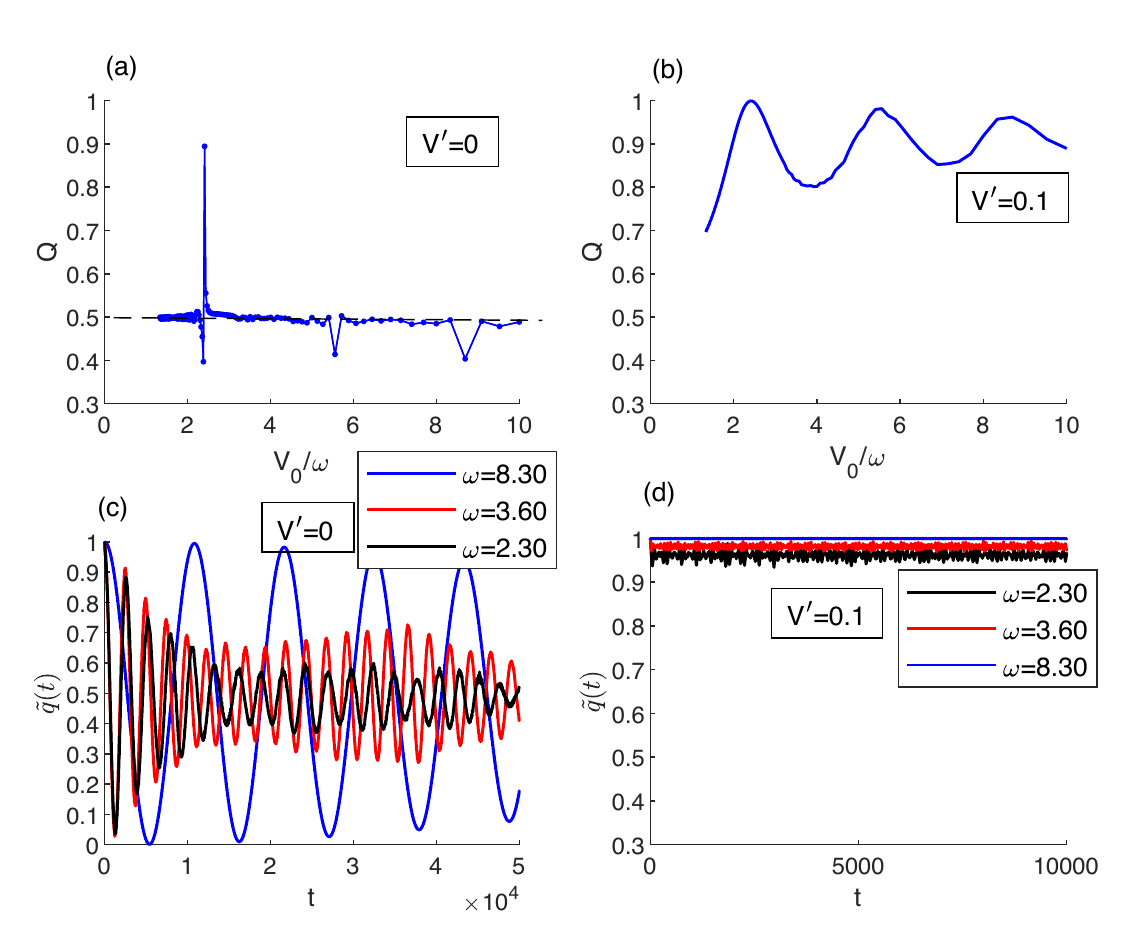}
\caption{\label{f-1p} One-particle sector: $Q(V_0/\omega)$ for $N=100$ (a)~$V'=0$, (b)~$V' = 0.1$; and $\tilde q(t)$ for specific values of $\omega$ for $N=100$ with  (c)~$V'=0 $, and (d)~$V' = 0.1$. In all the cases $V_0=20$.}
\end{figure*}

Figure \ref{f-1p}a  shows $Q$ (which is $q_{\bf k}( t)$ averaged over both ${\bf k}$ and $t$) as a function of the dimensionless parameter $V_0/\omega$. We have fixed $V_0=20$ and varied $\omega$, and the time evolution is calculated for a duration of $T=2000$. $Q(V_0/\omega)$ has a value close to $0.5$ (dashed line) for almost all values of $\omega$, except around three points where it shows some deviation from the constant value. In particular, around $\omega=8.3 ~(V_0/\omega = 2.4)$ the system appears to be freezing. For the three special frequencies ($\omega = 2.3,~3.6,$ and $8.3$), we have evolved the system for a longer time ($T=50000$); Fig. \ref{f-1p}c shows plots of corresponding $\tilde q(t)$. $Q$ obtained by averaging over the longer duration are, respectively, $0.48, ~0.49,$ and $0.49$, which are closer to the constant value of $0.5$ we obtained for other frequencies. Therefore, at large times, the system does not freeze at any value of the frequency.  

To gain a better understanding as to why there is no freezing at any frequency, as has been the case with two-band models such as the Ising model, and to find possible routes towards freezing, we next analyze the dynamics using a rotating wave approximation.

\subsubsection{\label{s-bgrwa} Rotating wave approximation} 

To implement the rotating wave approximation (RWA) \cite{AshJoh(07), SenSen(21)}, we first do the following time-dependent unitary transformation that takes us to a rotating frame (in the interaction picture):
\begin{align}
U &=\exp{-i\int_0^t H_V(t')dt'},
\label{e-ubg0}
\end{align}
where $H_V$ is the time-dependent part of $H_{\bf k}$:
\begin{align}
    H_V &= -\frac{V_{0}}{2 \omega}\cos{(\omega t)} ~\sigma^{z}\otimes I.
\end{align}
Substituting for $H_V$, we get
\begin{align}
U &=\exp{i\left(\frac{V_{0}}{2 \omega}\sin{(\omega t)}\right)(\sigma^{z}\otimes I)},
\label{e-ubg}
\end{align}
and then the effective Hamiltonian, $H'_{\bf k} = UH_{\bf k}U^\dagger + i \partial_t U U^\dagger$, is 
 \begin{align}
  H^{\prime}_{\bf k}= 
\begin{bmatrix}
  0 & -uf_{\bf k} & 0 & 0    \\\\
    -uf^*_{\bf k} & 0 & u'e^{2i\theta} & 0  \\\\
    0 & u'e^{-2i\theta} & 0 &  -uf_{\bf k}      \\\\
    0 & 0 & -uf^*_{\bf k} & 0  \\\\
 \end{bmatrix},
 \label{e-hrbg}
\end{align}
where $\theta=\left(V_{0}/2\omega\right)\sin{\omega t}$. 

Expanding $\exp[i(V_0/\omega)\sin ({\omega t})]$ in terms of $e^{in\omega t}$, $n\in \mathbb{Z}$, we get
\begin{align}
    \exp[i\left(\dfrac{V_0}{\omega}\right) \sin (\omega t)]  = \sum_{n=-\infty}^{\infty}  J_{n}\left(\dfrac{V_0}{\omega}\right) e^{i n \omega t},
    \label{e-bessel}
\end{align}
where $J_n(V_0/\omega)$ are the Bessel's functions of the first kind. In RWA, for large $\omega$, we retain only the $n=0$ term, neglecting all the faster-oscillating terms \cite{AshJoh(07)}. Then the Hamiltonian becomes
\begin{align}
  H^{\prime}_{\bf k}= 
\begin{bmatrix}
  0 & -\alpha_{\bf k}& 0 & 0    \\\\
    -\alpha^*_{\bf k} & 0 & \beta & 0  \\\\
    0 & \beta & 0 &  -\alpha_{\bf k}     \\\\
    0 & 0 & -\alpha^*_{\bf k} & 0  \\\\
 \end{bmatrix},
 \label{e-rwabg}
\end{align}
where $\alpha_{\bf k} = uf_{\bf k}$, and $\beta = u'J_{0}(V_{0}/\omega)$.

Evaluating $q_{\bf k}(t)$ [Eq. (\ref{e-qkt})], the probability for a particle with momentum ${\bf k}$ to remain in the initial state at time $t$, we obtain [see Eq. (\ref{ea-qkt1p})]
 \begin{align}
  q_{\bf k}( t) =&\dfrac{1}{4}\Big[\cos(\lambda_{1}t)+\cos(\lambda_{2}t)\Big]^2\nonumber\\
     &+16u^2 f_{\bf k}^2\left[\dfrac{\lambda_{1}}{N_{1}} \sin(\lambda_{1}t) +\dfrac{\lambda_{2}}{N_{2}} \sin(\lambda_{2}t)\right]^2,
  \label{e-qbg}
\end{align}
where \begin{align}
    \lambda_{1}&=-\dfrac{1}{2}\left(\beta+\sqrt{4\vert \alpha_{\bf k}\vert^2+\beta^2}\right),\\
    \lambda_{2}&=\dfrac{1}{2}\left(\beta-\sqrt{4\vert \alpha_{\bf k}\vert^2+\beta^2}\right),
\end{align}   
and  $N_{1}=8(\vert \alpha_{\bf k}\vert^{2}+\lambda_{1}^2),~N_{2}=8(\vert \alpha_{\bf k} \vert^{2}+\lambda_{2}^2).$
The time-average of $q_{\bf k}( t)$  is then [Eq. (\ref{ea-qbar1p})]
 \begin{align}
    \bar{q}_\textbf{k}=\dfrac{1}{4}+\dfrac{|\alpha_{\bf k}|^{2}}{4|\alpha_{\bf k}|^{2}+\beta^{2}}.
    \label{e-qbar1p}
\end{align}

The maximum of $\bar{q}_\textbf{k}$ is when $\beta = 0$. Thus, $\bar{q}_\textbf{k} \le 1/2$, for all ${\bf k}$. We can also obtain a lower bound by noting that $\bar{q}_\textbf{k}$ has its lowest value when $\beta^2$ takes its maximum value and $|\alpha_{\bf k}|$ its  minimum. From Eq. (\ref{e-bilmomh}) it follows that $|\alpha_{\bf k}|_{min} = u$, and $\beta_{max} = u'$ (since $J_0(V_0/\omega) \le 1$). With our choice of values for the parameters ($u=1,~u'=0.2$), we get $\bar{q}_\textbf{k} \ge 0.4975$. Putting the two bounds together,
\begin{align}
 0.4975 \le \bar{q}_\textbf{k} \le 0.5.
 \label{e-1pbound}
\end{align}
Thus, according to RWA, $\bar{q}_\textbf{k} \approx 0.5$ for all ${\bf k}$, independent of $V_0$ and $\omega$. Consequently, $Q(V_0/\omega)$, the average of $\bar{q}_\textbf{k}$ over ${\bf k}$, is also approximately 0.5 for all values of $V_0/\omega$; there is no freezing.

The RWA value of $Q(V_0/\omega) \approx 0.5$ is in good agreement with our numerical calculations (Fig. \ref{f-1p}a), except around the three specific values of $\omega$ we discussed earlier. The deviation of $Q$ from its RWA value for these frequencies can be understood as follows. In Eq. (\ref{e-qbg}) for $q_{\bf k}(t)$, there are terms of the form $\cos{(\lambda_1-\lambda_2)t}$ and $\sin{(\lambda_1-\lambda_2)t}$. The time-average of such terms over an interval $T$ will vanish if $T\gg (\lambda_1-\lambda_2)^{-1}$. However, as $J_0(V_0/\omega) \rightarrow 0$, $(\lambda_1-\lambda_2)^{-1} \rightarrow \infty$, therefore, for $\bar{q}_\textbf{k}$ to converge to its long-time average, the time over which the averaging is done should approach $\infty$. Consequently, around those values of $V_0/\omega$ for which $J_0(V_0/\omega)=0$, the convergence of $Q$ to its RWA value will be extremely slow. In Fig. \ref{f-1p}a,  the spikes in $Q(V_0/\omega)$ occur around $V_0/\omega = 2.30,~ 3.60$ and 8.30 whereas the zeroes of $J_0(V_0/\omega)$ are at $2.40,~ 5.52,$ and $8.65$. 

When $J_0(V_0/\omega) = 0$, and therefore $\beta = 0$, the initial state [Eq. (\ref{e-Gsbg})] becomes an eigenstate of the $H'_{\bf k}$ [Eq. (\ref{e-rwabg})] for all ${\bf k}$, which would then imply that the state is stationary and therefore $\bar{q}_{\bf k} = 1$. However, when $\beta = 0$, the initial state also becomes degenerate with the state
  \begin{align}
   \ket{\psi_{\bf k}^{\prime}}_{1p}=\dfrac{1}{\sqrt{2}}
   \begin{bmatrix}
     0\\
     0\\
     1\\
     \dfrac{\vert f_{\bf k}\vert}{f_{\bf k}}
   \end{bmatrix},
   \label{e-GsOp2}
  \end{align}
both having eigenvalue $-|uf_{\bf k}|$. Then, for arbitrarily small values of $\beta$, which couples these two degenerate states, there will be full oscillation between the two states. Therefore, we must take the limit $\beta\rightarrow 0$ of the general expression for $\bar q_{\bf k}$ [Eq. (\ref{e-qbar1p})] to get its physically correct value instead of directly putting $\beta = 0$ in the Hamiltonian. Taking the limit, we get
\begin{align}
    \lim_{\beta\rightarrow 0} \bar q_{\bf k} = \frac{1}{2},
    \label{e-qbarj0}
\end{align}
and therefore $Q=1/2$ as well. That is, there is no freezing even for those values of $\omega$ at which $\beta =0$.

It is the degeneracy in the rotating wave Hamiltonian that prevents the system from freezing even as $\beta \rightarrow 0$. A simple way to lift the degeneracy is to introduce a constant bias in driving, which we consider next. 

\subsubsection {\label{s2-driving wbia} Periodic driving with bias}

Adding a constant term, the potential becomes $V(t) = V' + V_{0} \cos (\omega t)$.  As before, going to the rotating frame via the transformation
\begin{align}
U &=\exp[i\left(\frac{V_{0}}{2 \omega}\sin{(\omega t)} + \frac{V't}{2}\right)(\sigma^{z}\otimes I)],
\label{e-uintbg}
\end{align}
and then applying the rotating wave approximation, we obtain the effective Hamiltonian to be
 \begin{align}
  \tilde H_{\textbf{k}}= 
\begin{bmatrix}
  0 & -\alpha_{\bf k} & 0 & 0    \\\\
    -\alpha^*_{\bf k} & 0 & \beta e^{iV't} & 0  \\\\
    0 & \beta e^{-iV't} & 0 &  -\alpha_{\bf k}      \\\\
    0 & 0 & -\alpha^*_{\bf k} & 0  \\\\
 \end{bmatrix}.
\end{align}
 The rotating wave Hamiltonian can be made time-independent by yet another transformation that takes $\ket{3} \rightarrow e^{-iV't}\ket{3}$, $\ket{4} \rightarrow e^{-iV't}\ket{4}$ and leaves $\ket{1}$ and $\ket{2}$ invariant. The resultant effective Hamiltonian is then    
\begin{align}
  H''_{\textbf{k}}= 
\begin{bmatrix}
  0 & -\alpha_{\bf k} & 0 & 0    \\\\
    -\alpha^*_{\bf k} & 0 & \beta & 0  \\\\
    0 & \beta & -V^{\prime} &  -\alpha_{\bf k}      \\\\
    0 & 0 & -\alpha^*_{\bf k} & -V^{\prime}  \\\\
 \end{bmatrix}
 \label{e-efhbg}
\end{align}
When $\beta = 0$ (i.e., when $J_0(V_0/\omega) = 0$), the initial state [Eq. (\ref{e-Gsbg})] is again a stationary state, but in the presence of $V'$ the corresponding eigenvalue is no longer degenerate. Then, $\bar{q}_{\bf k} \rightarrow 1$ as $J_0(V_0/\omega) \rightarrow 0$, and the system freezes.

For $V' = 0.1$, we have numerically calculated the dynamical freezing factor $Q(V_0/\omega)$ by varying $\omega$, keeping $V_0$ fixed at 20 (Fig. \ref{f-1p}b).  The system freezes almost completely at $V_0/\omega=$ 2.394, 5.509, and 8.620. These values of $V_0/\omega$ are in good agreement with the three zeroes of $J_0(V_0/\omega)$, which are at $V_0/\omega =$ 2.405, 5.520, and 8.654, respectively. In Fig. \ref{f-1p}d, we have plotted the response function $\tilde{q}(t)$  for those values of $\omega$ at which $Q(V_0/\omega)$ peaks; in all cases $\tilde{q}(t)\approx 1$ at all times.

\subsection{At half-filling}

\begin{figure*}
\includegraphics[width=0.8\textwidth]{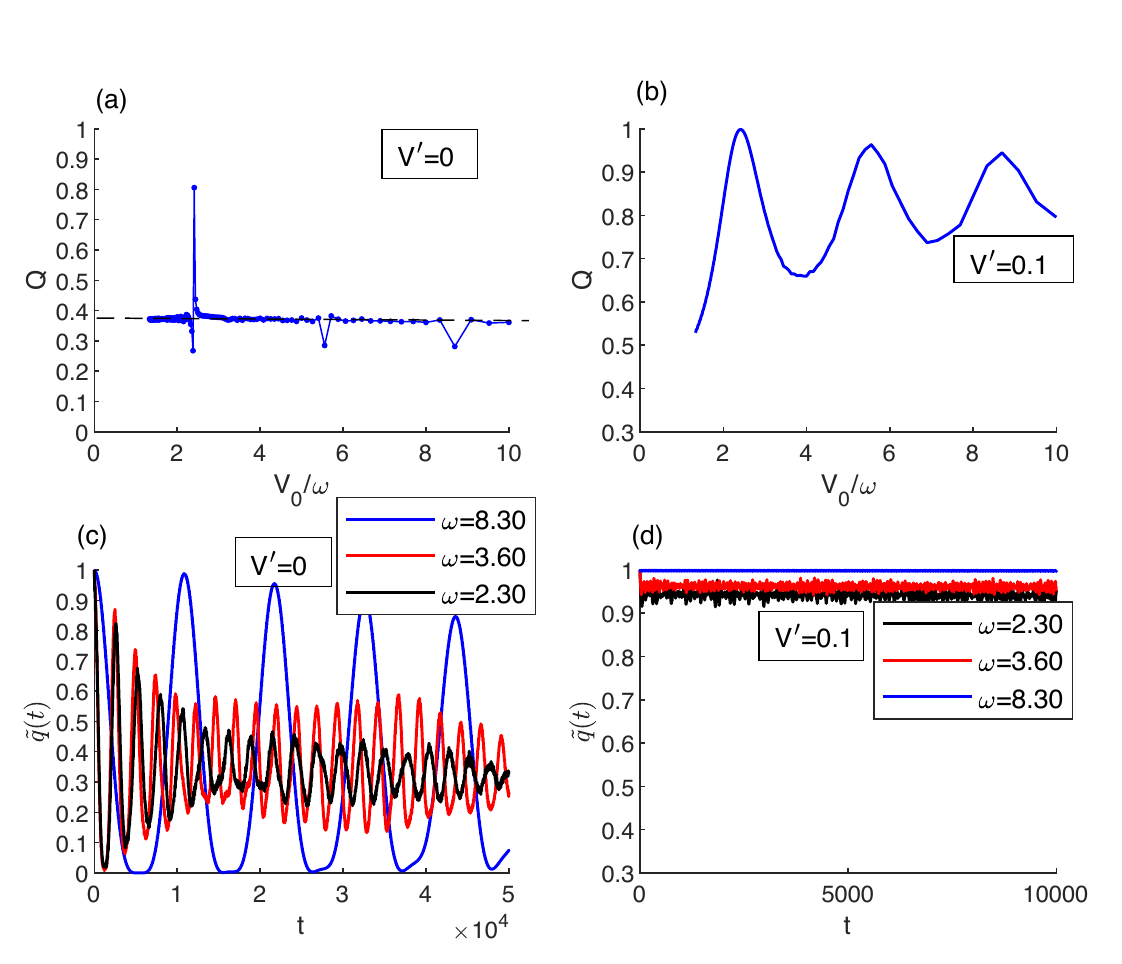}
\caption{\label{f-2p} Two-particle sector: $Q(V_0/\omega)$ for $N=100$ (a)~$V'=0$, (b)~$V' = 0.1$; and $\tilde q(t)$ for various values of $\omega$ for $N=100$  (c)~$V'=0 $, (d)~$V' = 0.1$. In all the cases, $V_0=20$. In (a), $Q(V_0/\omega)$ appreciably deviates from $0.375$ around $V_{0}/\omega \approx 2.4,~5.6,$ and $8.7$. }
\end{figure*}

Next, we consider the dynamics of the ground state at half-filling, which, as mentioned earlier, lies in the two-particle sector. We can compute the two-particle Hamiltonian $H_{{\bf k},2p}$ directly from its one-particle counterpart $H_{\bf k}$ [Eq. (\ref{e-bgham})]. Choosing the two-particle basis states as follows,
\begin{align}
\begin{split}
    \ket{1}_{2p} &= \frac{1}{\sqrt{2}} \left( \ket{1}_{1p} \otimes \ket{2}_{1p} - \ket{2}_{1p} \otimes \ket{1}_{1p} \right),\\
    \ket{2}_{2p} &= \frac{1}{\sqrt{2}} \left( \ket{1}_{1p} \otimes \ket{3}_{1p} - \ket{3}_{1p} \otimes \ket{1}_{1p} \right),\\
    \ket{3}_{2p} &= \frac{1}{\sqrt{2}} \left( \ket{1}_{1p} \otimes \ket{4}_{1p} - \ket{4}_{1p} \otimes \ket{1}_{1p} \right),\\
    \ket{4}_{2p} &= \frac{1}{\sqrt{2}} \left( \ket{2}_{1p} \otimes \ket{3}_{1p} - \ket{3}_{1p} \otimes \ket{2}_{1p} \right),\\
    \ket{5}_{2p} &= \frac{1}{\sqrt{2}} \left( \ket{2}_{1p} \otimes \ket{4}_{1p} - \ket{4}_{1p} \otimes \ket{2}_{1p} \right),\\
    \ket{6}_{2p} &= \frac{1}{\sqrt{2}} \left( \ket{3}_{1p} \otimes \ket{4}_{1p} - \ket{4}_{1p} \otimes \ket{3}_{1p} \right),
    \end{split}
    \label{e-2pbasis}
    \end{align}
where $\ket{j}_{1p}$ ($j=1~\mathrm{to}~4$) are the canonical one-particle basis states, we obtain
\begin{align}
H_{{\bf k},2p} &=
\begin{bmatrix}
    -V & u' & 0 & 0 & 0 & 0 \\\\
    u' & 0 & -\alpha_{\bf k} & -\alpha_{\bf k} & 0 & 0 \\\\
    0 & -\alpha^*_{\bf k} & 0 & 0 & -\alpha_{\bf k} & 0 \\\\
    0 & -\alpha^*_{\bf k} & 0 & 0 & -\alpha_{\bf k} & 0 \\\\
    0 & 0 & \alpha^*_{\bf k} &\alpha^*_{\bf k} & 0 & u'\\\\
    0 & 0 & 0 & 0 & u'& V
\end{bmatrix},
\label{e-2pham}
\end{align}
where $\alpha_{\bf k}= u f_{\bf k}$, as defined earlier. 

\subsubsection{Numerical analysis}

As in the one-particle case, we numerically compute $q_{\bf k}(t)$, the probability of the two-particle state in the sector with momentum ${\bf k}$ to remain in the initial ground state at a later time $t$. For $V_0\gg 1$, the initial ground state is $(1, 0, 0, 0, 0, 0)$. We first consider the case with $V'=0$.

Figure \ref{f-2p}a shows $Q(V_0/\omega)$ obtained by averaging $\tilde{q}(t)$ over a duration $T=2000$, keeping $V_0 = 20$ and varying $\omega$. As we anticipate from the one-particle case, there is no freezing at any value of $\omega$, with $Q$ having a nearly constant value, which is close to $0.375$ (dashed line). Moreover, $Q$ again deviates from the constant value around three specific frequencies; for $\omega= 2.3, 3.6$ and $8.3$, $Q$ is, respectively, 0.281, 0.285 and 0.806. As before, this deviation disappears when the averaging is done over longer times. Figure \ref{f-2p}c shows $\tilde{q}(t)$ corresponding to the above three values of $\omega$. When the time-averaging is done over $T=50000$, the respective values of $Q$ become 0.3430, 0.3461, and 0.3391, which are closer to the constant value of $0.375$ we obtained for generic values of frequency. 

Next, we consider the case with bias. We have plotted $Q(V_0/\omega)$ in Fig. \ref{f-2p}b. The peak values of $Q$ are very close to one, showing that the system is almost completely frozen at those frequencies. In Fig. \ref{f-2p}d we have shown $\tilde q(t)$ for the three frequencies that correspond to the peaks in $Q(V_0/\omega)$. In all three cases, $\tilde q(t)$  remains close to one at all times.

\subsubsection{\label{sec-rwa2p} Rotating wave approximation}
We now generalize the RWA calculations to the two-particle sector. For driving without bias, i.e., $V(t) = V_0 \cos{\omega t}$, the rotating wave Hamiltonian becomes
\begin{align}
H'_{{\bf k},2p} &=
\begin{bmatrix}
    0 & \beta & 0 & 0 & 0 & 0 \\\\
    \beta & 0 & -\alpha_{\bf k} & -\alpha_{\bf k} & 0 & 0 \\\\
    0 & -\alpha^*_{\bf k} & 0 & 0 & -\alpha_{\bf k} & 0 \\\\
    0 & -\alpha^*_{\bf k} & 0 & 0 & -\alpha_{\bf k} & 0 \\\\
    0 & 0 & \alpha^*_{\bf k} &\alpha^*_{\bf k} & 0 & \beta\\\\
    0 & 0 & 0 & 0 & \beta& 0
\end{bmatrix},
\label{e-2phamrw}
\end{align}
where $\beta = u'J_{0}\left(\dfrac{V_{0}}{\omega}\right)$. Then the probability to remain in the initial state is [Eq. (\ref{ea-2p-qkt})]
\begin{align}
    q_{\bf k}(t)=\Bigg[\dfrac{1}{2}\cos {\beta t} +& \dfrac{1}{N_1^{2}}\Big[4\vert \alpha_{\bf k} \vert^{2} \nonumber \\
    &+ {\beta^2} \cos \left({\sqrt{4\vert \alpha_{\bf k}\vert^{2}+\beta^{2}} t}\right)\Big]\Bigg]^{2}, \label{e-2p-qkt}
\end{align}
and the long-time average of $q_{\bf k}(t)$ becomes [Eq. (\ref{ea-2p-qbar})]
\begin{align}
\bar q_{\bf k} &=\dfrac{1}{8}+\dfrac{1}{8}\left[\dfrac{32 + x_{\bf k}^{2}}{(4+x_{\bf k})^2} \right],
  \label{e-2pqbar}
\end{align}
where $x_{\bf k}= \left(\beta^2/|\alpha_{\bf k}|^2\right)$. $\bar q_{\bf k}$ is a function only of $x_{\bf k}$ and has a minimum value of $5/24$ at $x_{\bf k}=8$. Further, $\bar q_{\bf k}\rightarrow 1/4$ as ${x_{\bf k}\rightarrow \infty}$, and $\bar q_{\bf k}$ has its maximum value of $3/8$ at $x_{\bf k}=0$. [Here we note that when $f_{\bf k}$ is exactly zero, directly from the Hamiltonian we obtain $\bar q_{\bf k}=1/2$, different from the value of $1/4$ obtained by taking the limit $f_{\bf k}\rightarrow 0$. This is because arbitrarily small $f_{\bf k}$ introduces nonzero matrix element between the initial state and the state degenerate with it.]

For our choice of parameter values ($u=1$ and $u'=0.2$), $\beta^2 \le 0.04$ (since $|J_0(V_0/\omega)|\le 1$). Therefore, $x_{\bf k} \ll 1$ unless $|f_{\bf k}|^2 \ll 0.04$. Now $|f_{\bf k}| = 0$ when ${\bf k} = {\bf K}$ or ${\bf K}'$ [Eq. (\ref{e-diracpts})]. Expanding  $|f_{\bf k}|$ around these points, it follows that the condition $x_{\bf k} \ll 1$ is violated only within circles of radius $|a_0k| \sim 0.133$ around the two gapless points. Consequently, for any frequency $\omega$, $\bar q_{\bf k}$ attains a value close to $3/8$ (its maximum, which corresponds to $x_{\bf k}=0$) for most values of ${\bf k}$. Then $Q$ is also approximately $3/8$, independent of $\omega$. The RWA calculation of $Q(V_0/\omega)$ is again in good agreement with the numerical values. In Fig. \ref{f-2p}a, the dashed line  corresponds to $Q(V_0/\omega)=0.375$, its RWA value. 

In the two-particle case, the deviation of $Q$ from the constant RWA value at the three specific frequencies is due to the slowing down of dynamics as $J_0(V_0/\omega)$ (and correspondingly $\beta$) approaches zero, as can be seen from Eq. (\ref{e-2p-qkt}).

For nonzero bias, i.e., $V'\neq 0$, the rotating wave Hamiltonian becomes
\begin{align}
H_{{\bf k},2p}''&=
\begin{bmatrix}
    0 &\beta & 0 & 0 & 0 & 0 \\\\
   \beta & -V^{\prime} & -\alpha_{\bf k} & -\alpha_{\bf k} & 0 & 0 \\\\
    0 &-\alpha^{*}_{\bf k} & -V^{\prime} & 0 & -\alpha_{\bf k} & 0 \\\\
    0 &-\alpha^{*}_{\bf k} & 0 & -V^{\prime} & -\alpha_{\bf k} & 0 \\\\
    0 & 0 &-\alpha^{*}_{\bf k} &\alpha^{*}_{\bf k} & -V^{\prime} &\beta\\\\
    0 & 0 & 0 & 0 &\beta& -2V^{\prime}
\end{bmatrix}
\end{align}
When $J_0(V_0/\omega) = 0$, and correspondingly $\tilde\beta_{\bf k} = 0$, as in the one-particle case, the initial state becomes an eigenstate of the rotating wave Hamiltonian for all ${\bf k}$ and the system freezes. Numerically, the peaks of $Q(V_0/\omega)$ occur at $V_0/\omega=$ 2.407, 5.525, and 8.639 (see Fig. \ref{f-2p}b), which again compares well with the zeroes of $J_0(V_0/\omega)$: 2.404, 5.520, and 8.654, respectively.

\subsection{Switching of response}

\begin{figure*}
\includegraphics[width=0.8\textwidth]{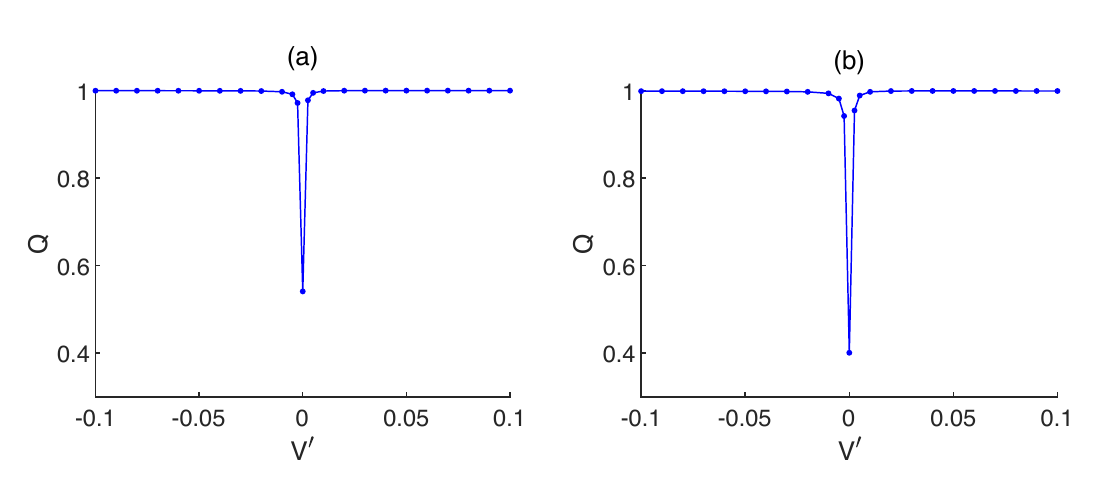}
\caption{\label{f-switch} $Q(V')$ for $N=100$, $V_0=20$ and $\omega = 8.3$: (a) one-particle sector, (b) two-particle sector.}
\end{figure*} 

We have shown above that freezing occurs at special values of $V_0/\omega$, but only when the driving has a nonzero bias. There is no freezing (for any combination of frequency and amplitude) when $V'=0$. In Fig. \ref{f-switch}, the quantity $Q$ is plotted against $V'$ for $V_0=20$ and $\omega = 8.3$, a combination of values at which freezing occurs for non-zero $V'$. $Q$ drops sharply to its non-freezing value as $V'$ approaches zero. Similar switching of response by tuning a parameter has been found earlier in the transverse-field $X$-$Y$ model \cite{DasMoe(12)}. 

\section{\label{s-oplat} Realization via an optical lattice}

In this section, we give a proposal for an optical lattice realization of the bilayer Hamiltonian using cold atoms, where the driving can be implemented via lattice shaking.  By oscillating the optical lattice potential, the atoms experience an effective periodic force \cite{Eck(17)}. Dynamical localization via lattice shaking has been achieved in a one-dimensional optical lattice \cite{LigSia(07)}. Optical lattice shaking has also been used to dynamically control quantum phase transition between Mott insulator to superfluid \cite{ZenLig(09)} and to study coherent resonant ac-induced tunneling \cite{IvaSch(08)}, to name a few other applications.

Optical lattices are created by counter-propagating laser beams, appropriately aligned along different directions, which create a periodic potential that can trap atoms at its minima \cite{GriWei(99), BloDal(08)}. Atoms can then hop between neighboring potential minima, with the hopping amplitudes determined by the depth of the potential wells. When the potential wells are sufficiently deep, only hopping between the nearest neighboring sites needs to be considered.

Since the two layers are identical in the $AA$-stacked bilayer \cite{RozSbo(16)}, its implementation is more straightforward compared to the $AB$-stacked bilayer, and therefore we consider the former. 

A honeycomb optical lattice is created by three counter-propagating laser beams, resulting in the following potential \cite{ShiWan(07),ShaShi(08)} 
\begin{align}
    V(x,y)=\sum_{j=1,2,3}V_{0}\sin^{2}\left[{k}_{L}(x\cos{\theta_{j}}+y\sin{\theta_{j}})\right],
\label{e-Vxy}
\end{align}
where $\theta_{1}=\pi/3$,~$\theta_{2}=2\pi/3$, $\theta_{3}=0$, and $\textbf{k}_{L}$ is the optical wave vector in the $xy$ plane. 

\begin{figure}
    \includegraphics[width=0.4\textwidth]{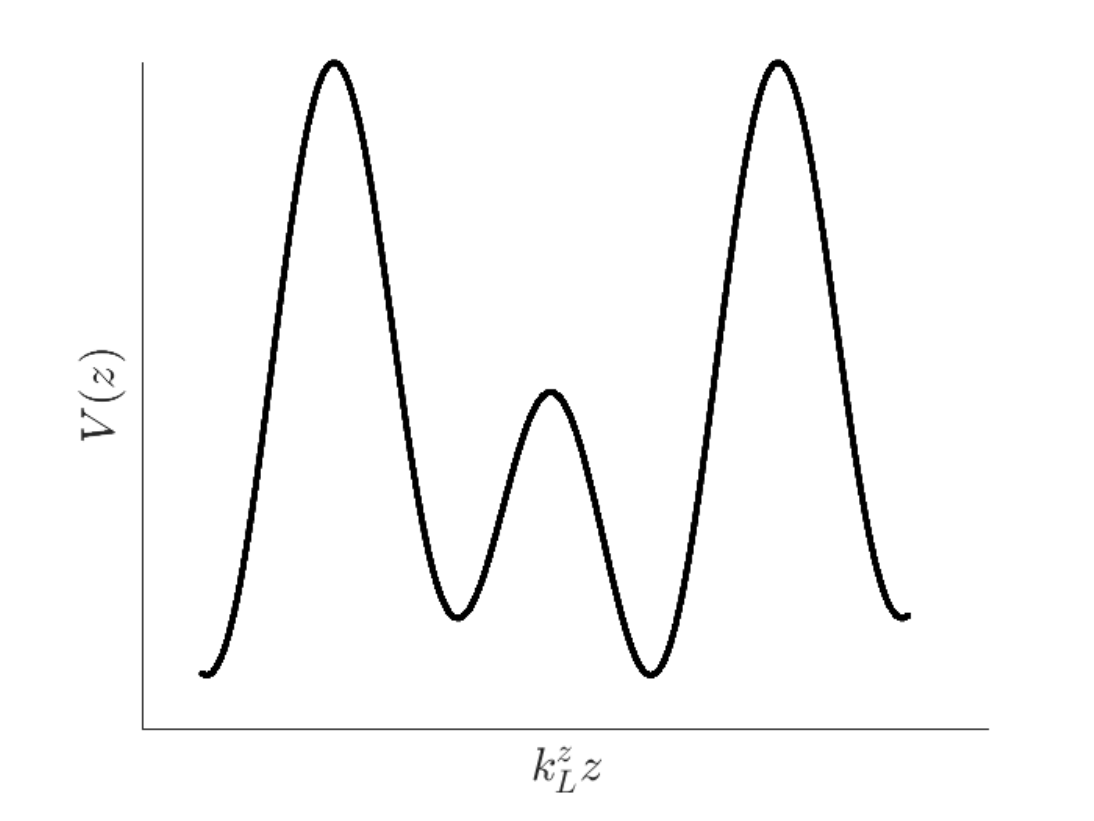}
\caption{Asymmetric double-well potential along the $z$-direction}
\label{f-doublewell}
\end{figure}

To create the bilayer, two sets of counter-propagating laser beams in the $z$-direction with wave vector components  $k_L^z$ and $2k_L^z$ are introduced, which results in a confining double-well potential in the $z$-direction \cite{Wuhe(13)}:
\begin{align}
    V(z)=V_{L}\sin^{2}(k_{L}^{z} z)-V_{S}\sin^{2}(2 k_{L}^{z} z + \phi_0),
    \label{e-Vz}
\end{align}
where $V_{L}$ and $V_{S}$ are the amplitudes of the long and short lasers. The phase difference $\phi_0$ between the two terms creates an asymmetric double-well potential \cite{JulEig(13)}, as shown in Fig. \ref{f-doublewell}. By appropriately choosing the amplitudes $V_L$ and $V_S$ the atoms can be confined to a single double-well along the $z$-direction. Then, the minima of the sum of two potentials $V(x,y)$ and $V(z)$ in Eqs. (\ref{e-Vxy}) and (\ref{e-Vz}) form an $AA$-stacked bilayer honeycomb lattice.

Allowing for only nearest neighbor hopping, the momentum-space Hamiltonian for $AA$-stacking is
\begin{align}
  H_{\textbf{k}}= 
  \begin{bmatrix}
-\dfrac{V'}{2} & -u f_{\bf k} & u' & 0 \\
-u f^*_{\bf k} & -\dfrac{V'}{2} & 0 & u'\\
u' & 0 & \dfrac{V'}{2} & -u f_{\bf k}   \\
0 & u' &-u f^*_{\bf k}  & \dfrac{V'}{2} \\ 
\end{bmatrix},
\label{e-bghamAA}
 \end{align}
 where, as in the case of $AB$-stacking,
 \begin{align}
     f_{\bf k}=\exp(-ia_{0}k_{x})\left[1+2\exp(\dfrac{3ia_{0}k_{x}}{2})\cos(\dfrac{\sqrt{3}a_{0}k_{y}}{2})\right].
     \label{e-bilmomhAA}
     \end{align}
$2V'$ is the potential difference between the two layers arising from the asymmetry. The only difference between the above Hamiltonian and that of $AB$-stacking [Eq. (\ref{e-bgham})] is in the position of $u'$.

\subsection{Periodic driving via lattice shaking}

We now periodically drive the system by shaking the lattice. We show that shaking the lattice in the direction normal to the plane of the lattice is exactly equivalent to driving the interlayer potential difference. 

Lattice shaking can be achieved by introducing a small difference in frequency $\Delta \nu$ between the two counter-propagating laser beams and then varying $\Delta \nu$ sinusoidally: $\Delta \nu = \Delta \nu_{max} \sin \omega t$ \cite{LigSia(07)}. Here  we shake the lattice along the $z$-direction (i.e., perpendicular to the plane of the lattice). Then the lattice moves with velocity  \mbox{${\bf v}(t) = a_0 \Delta \nu_{max} \sin \omega t ~\hat z$}, which results in the atoms experiencing the periodic force 
\begin{align}
    \textbf{F(t)} = m\omega a_0 \Delta \nu_{max} \cos \omega t ~\hat{z}.
\end{align}
Here $m$ is the mass of an atom and $a_0$ the lattice spacing. In the lattice frame of reference, this results in the following additional term in the Hamiltonian \cite{Eck(17)}:
\begin{align}
    V(t) &= \sum_j w_j(t) \hat n_j,
\end{align}
where $j$ denotes a lattice site, $\hat n_j$ is the corresponding number operator, and 
\begin{align}
     w_{j}(t)=-\textbf{r}_{j}\cdot \textbf{F}(t),
 \end{align}
 where ${\bf r}_j$ is the lattice vector at site $j$. Applying the gauge transformation
\begin{align}
   U(t)=\exp{i \sum_{j}\chi_{j}(t)\hat{n}_{j}},
\end{align}
where,
\begin{align}
 \chi_{j}(t)=-\int_{0}^{t}w_{j}(t')dt',  
\end{align}
the hopping terms transform as follows:
\begin{align}
    a_j^{\dagger}a_{j'} \rightarrow e^{i(\chi_j-\chi_{j'})} a_j^{\dagger}a_{j'}.
\end{align}
Since the driving force is in the $z$-direction, only the inter-layer hopping terms transform nontrivially. Eventually, we obtain the momentum space Hamiltonian as
\begin{align}
  H_{\textbf{k}}= 
  \begin{bmatrix}
-\dfrac{V'}{2} & -u f_{\bf k} & \tilde u' & 0 \\
-u f^*_{\bf k} & -\dfrac{V'}{2} & 0 & \tilde u'\\
{\tilde u'}{}^* & 0 & \dfrac{V'}{2} & -u f_{\bf k}   \\
0 & {\tilde u'}{}^*  &-u f^*_{\bf k}  & \dfrac{V'}{2} \\ 
\end{bmatrix},
\label{e-bghamAA1}
 \end{align}
where $\tilde u'=u'e^{i(K/\omega)\sin{\omega t}}$, and $K = m\omega a_0^2 \Delta \nu_{max}$.
The above Hamiltonian has the same form as the one we obtained earlier by driving the voltage difference between the layers, which for $AA$-stacking [after modifying Eq. (\ref{e-hrbg})] is
\begin{align}
  H^{\prime}_{\bf k}= 
\begin{bmatrix}
  -\dfrac{V'}{2} & -uf_{\bf k} & u'e^{2i\theta} & 0    \\\\
    -uf^*_{\bf k} & -\dfrac{V'}{2} & 0 & u'e^{2i\theta}  \\\\
    u'e^{-2i\theta} & 0 & \dfrac{V'}{2} &  -uf_{\bf k}      \\\\
    0 & u'e^{-2i\theta} & -uf^*_{\bf k} & \dfrac{V'}{2}  \\\\
 \end{bmatrix},
 \label{e-hamAAshake}
\end{align}
where $\theta=\left(V_{0}/2\omega\right)\sin{\omega t}$, and $V'$ is the constant bias. The two Hamiltonians are identical when $V_0=K$. In other words, driving the interlayer potential difference is mathematically equivalent to shaking the lattice in the $z$-direction. 

\subsubsection{State preparation}
Our initial state is the ground state corresponding to large $V_0$. In order to replicate this situation in the optical lattice, we start with a strongly asymmetric potential $V(z)$ by appropriately choosing the phase difference $\phi_0$ in Eq. (\ref{e-Vz}). The ultracold atoms are then loaded into the layer at the lower potential. If we now start shaking the lattice keeping the asymmetry on, then the system will freeze at frequencies for which $J_0(K/\omega)=0$. On the other hand, if we switch off the asymmetry before starting the lattice shaking, there is no freezing at any $\omega$.

\vspace{1in}

\section{\label{s-sum} Summary and discussion}

We have studied the response of bilayer graphene to  harmonically driving the interlayer potential difference. When the driving is unbiased, i.e., when the induced potential varies symmetrically about zero, the system does not freeze for any combination of the driving amplitude and frequency. Using a rotating wave analysis, we have shown that for freezing to occur, a ground state degeneracy in the rotating frame should be lifted, which we achieve by introducing a bias to the driving. Then the system freezes almost absolutely for certain values of the ratio between the amplitude and frequency of driving. We support our numerical results with analytical calculations based on the rotating wave approximation. 

We have put forth a proposal to realize the bilayer system using ultracold atoms in an optical lattice. 
We have further shown that driving the interlayer potential can be simulated by shaking the lattice in the direction normal to the lattice plane. Dynamical localization via lattice shaking in a one-dimensional system has already been demonstrated in the lab \cite{LigSia(07)}, showing that the parameter regime in which dynamical freezing occurs is achievable.

\appendix

\section{\label{s-app} RWA calculation of $\bar q _{\bf k}$ for $V'=0$}

\subsection{One-particle sector}
The rotating wave Hamiltonian is
\begin{align}
  H^{\prime}_\textbf{k}= 
\begin{bmatrix}
  0 & -\alpha_{\bf k} & 0 & 0    \\\\
    -\alpha_{\bf k}^{*} & 0 & \beta & 0  \\\\
    0 & \beta & 0 &  -\alpha_{\bf k}     \\\\
    0 & 0 & -\alpha_{\bf k}^{*} & 0  \\\\
 \end{bmatrix},
\end{align}
where $\alpha_{\bf k}=uf_\textbf{k}$ and $\beta=u'J_{0}\left(\dfrac{V_{0}}{\omega}\right)$.
The eigenvalues of $H'_{\bf k}$ are 
 \begin{align}
 \begin{split}
    \lambda_{1} &=\dfrac{1}{2}\left(-\beta-\sqrt{4\vert \alpha_{\bf k}\vert^2+\beta^2}\right),\\
    \lambda_{2} &= \dfrac{1}{2}\left(\beta-\sqrt{4\vert \alpha_{\bf k}\vert^2+\beta^2}\right), \\
    \lambda_{3} &=\dfrac{1}{2}\left(-\beta+\sqrt{4\vert \alpha_{\bf k}\vert^2+\beta^2}\right),\\
    \lambda_{4} &= \dfrac{1}{2}\left(\beta+\sqrt{4\vert \alpha_{\bf k}\vert^2+\beta^2}\right), 
\label{e-1peigenvalues}
\end{split}
\end{align}
and the corresponding eigenbras are, respectively,
\begin{align}
\begin{split}
 \bra{\lambda_1} &= \dfrac{1}{{N_{1}}}
 \begin{bmatrix}
      -\alpha_{\bf k} & \lambda_{1} & -\lambda_{1} & \alpha_{\bf k}^{*}
    \end{bmatrix}, \\
\bra{\lambda_2} &= \dfrac{1}{{N_{2}}}
 \begin{bmatrix}
      \alpha_{\bf k}^* & -\lambda_{2} & -\lambda_{2} & \alpha_{\bf k}
    \end{bmatrix}, \\
\bra{\lambda_3} &=\dfrac{1}{{N_{2}}}
 \begin{bmatrix}
      -\alpha_{\bf k}^* & -\lambda_{2} & -\lambda_{2} & \alpha_{\bf k}
    \end{bmatrix}, \\
    \bra{\lambda_4} &=\dfrac{1}{{N_{1}}}
 \begin{bmatrix}
      \alpha_{\bf k}^* & \lambda_{1} & \lambda_{1} & \alpha_{\bf k}
    \end{bmatrix},
    \end{split}
  \end{align}
where $N_{1}=\sqrt{2\left(\vert \alpha_{\bf k} \vert^{2}+\lambda_{1}^2\right)}$ and  $N_{2}=\sqrt{2\left(\vert \alpha_{\bf k} \vert^{2}+\lambda_{2}^2\right)}$.
 
Writing the initial state $\ket{\psi_{\bf {k}}(0)}$ in terms of the eigenkets, we get 
 \begin{align}
    \ket{\psi_{{\bf k}}(0)}= \sum_{n=1}^4 C_{n}(0)\ket{\lambda_n},
\end{align}
where,
\begin{align}
\begin{split}
     C_{1}(0)&= \dfrac{-1}{\sqrt{2}N_{1}}\left[\alpha_{\bf k}^{*}-\lambda_{1}\dfrac{\vert f_{\bf k}\vert}{f_{\bf k}}\right],\\
      C_2(0) &= \dfrac{1}{\sqrt{2}N_{2}}\left[\alpha_{\bf k}^{*}-\lambda_{2}\dfrac{\vert f_{\bf k}\vert}{f_{\bf k}}\right],\\
  C_{3}(0) &= \dfrac{-1}{\sqrt{2}N_{2}}\left[\alpha_{\bf k}^{*}+\lambda_{2}\dfrac{\vert f_{\bf k}\vert}{f_{\bf k}}\right],\\
      C_{4}(0)&= \dfrac{1}{\sqrt{2}N_{1}}\left[\alpha_{\bf k}^{*}+\lambda_{1}\dfrac{\vert f_{\bf k}\vert}{f_{\bf k}}\right].
      \end{split}
  \end{align} 
 Then,
 \begin{align}
     q_{\bf k}( t)  &=|{\bra{\psi_{\bf k}(0)}\ket{\psi_{\bf k}(t)}}|^2 \nonumber\\ 
    &= \dfrac{1}{4}\left(\cos(\lambda_{1}t)+\cos(\lambda_{2}t)\right)^{2}\nonumber\\
  &~+64 \vert \alpha_{\bf k} \vert^{2}\left(\dfrac{\lambda_{1}}{N_{1}} \sin(\lambda_{1}t) +\dfrac{\lambda_{2}}{N_{2}} \sin(\lambda_{2}t)\right)^{2}. \label{ea-qkt1p}
  \end{align}
Finally, taking the time-average, we obtain (for $\lambda_1 \neq \lambda_2$)
 \begin{align}
    \bar{q}_\textbf{k}=\lim_{T\rightarrow \infty}\frac{1}{T}\int_0^T q_{\bf k}( t) dt = \dfrac{1}{4}+\dfrac{|\alpha_{\bf k}|^{2}}{4|\alpha_{\bf k}|^{2}+\beta^{2}}.
    \label{ea-qbar1p}
\end{align}
 
\subsection{Two-particle sector}

In the two-particle sector, the rotating wave Hamiltonian is
\begin{align}
H'_{{\bf k},2p}&=
\begin{bmatrix}
    0 &\beta & 0 & 0 & 0 & 0 \\\\
   \beta & 0 & -\alpha_{\bf k} & -\alpha_{\bf k} & 0 & 0 \\\\
    0 &-\alpha^{*}_{\bf k} & 0 & 0 & -\alpha_{\bf k} & 0 \\\\
    0 &-\alpha^{*}_{\bf k}& 0 & 0 &-\alpha_{\bf k} & 0 \\\\
    0 & 0 & -\alpha^{*}_{\bf k} &-\alpha^{*}_{\bf k} & 0 &\beta\\\\
    0 & 0 & 0 & 0 &\beta& 0
\end{bmatrix}.
\end{align}
Eigenvalues of $H'_{{\bf k},2p}$ can be written in terms of the eigenvalues of the one-particle Hamiltonian [Eqs. (\ref{e-1peigenvalues})]:
\begin{align}
\begin{split}
    \mu_{1} &=\lambda_{1}+\lambda_{4}=0,\\
    \mu_{2}&=\lambda_{2}+\lambda_{3}=0,\\
    \mu_{3}&=\lambda_{1}+\lambda_{3}=-\beta,\\
    \mu_{4}&=\lambda_{2}+\lambda_{4}=\beta,\\
    \mu_{5}&=\lambda_{1}+\lambda_{2}=-\sqrt{4\vert \alpha_{\bf k}\vert^{2}+\beta^{2}},\\
    \mu_{6}&=\lambda_{3}+\lambda_{4}=\sqrt{4\vert \alpha_{\bf k}\vert^{2}+\beta^{2}},
    \label{e-2peigenvalues}
\end{split}
\end{align}
and the corresponding eigenbras are
\begin{align*}
\bra{\mu_{1}}&=\dfrac{1}{N_1}
    \begin{bmatrix}
     {2}\alpha_{\bf k}^*& 0 &\beta&\beta&0& 2\alpha_{\bf k}  
    \end{bmatrix},\\
    \bra{\mu_{2}}&=\dfrac{1}{N_2}
    \begin{bmatrix}
     {0}& {0}&\beta&-\beta&0&0   
    \end{bmatrix},\\
    \bra{\mu_{3}}&=\dfrac{1}{2}
    \begin{bmatrix}
    -1& 1 & 0&0&-\dfrac{\alpha_{\bf k}}{\alpha^*_{\bf k}}& \dfrac{\alpha_{\bf k}}{\alpha^*_{\bf k}}
    \end{bmatrix},\\
    \bra{\mu_{4}}&=\dfrac{1}{2}
\begin{bmatrix}
    1& 1 & 0&0&-\dfrac{\alpha_{\bf k}}{\alpha^*_{\bf k}}& -\dfrac{\alpha_{\bf k}}{\alpha^*_{\bf k}}
\end{bmatrix},
\end{align*}
\vspace{-0.5cm}
\begin{align*}
\bra{\mu_{5}}&=\dfrac{1}{\sqrt{2}N_1}
    \begin{bmatrix}
     -{\beta}&\mu_{6} &2\alpha_{\bf k}&2\alpha_{\bf k}&\dfrac{\alpha_{\bf k}\mu_{6}}{\alpha^*_{\bf k}}&  -\dfrac{\alpha_{\bf k}{\beta}}{\alpha^*_{\bf k}}
    \end{bmatrix},\\
    \bra{\mu_{6}}&=\dfrac{1}{\sqrt{2}N_1}
    \begin{bmatrix}
     {\beta}&\mu_{6} &-2\alpha_{\bf k}&-2\alpha_{\bf k}&\dfrac{\alpha_{\bf k}\mu_{6}}{\alpha^*_{\bf k}}&  \dfrac{\alpha_{\bf k}{\beta}}{\alpha^*_{\bf k}}
     \end{bmatrix}
    \end{align*}
    where $N_1 =\sqrt{8\vert \alpha_{\bf k} \vert^{2}+2\beta^{2}}$ and $N_2 =\sqrt{2}\beta$. The initial state $\ket{\psi_{\bf k}(0)}_{2p}=\ket{1}$, therefore, 
\begin{align*}
    {\bra{\psi_{\bf k}(0)} \ket{\psi_{\bf k}(t)}} &= \sum_j |x_{1,j}|^2 e^{-i\mu_j t} \\
    &= \dfrac{1}{2}\cos {\mu_3 t}+\dfrac{1}{N_1^{2}}\left[4\vert \alpha_{\bf k} \vert^{2} + {\beta^2} \cos {\mu_5 t}\right].
\end{align*}
Then,
\begin{align}
    q_{\bf k}(t)&= \left|{\bra{\psi_{\bf k}(0)}\ket{\psi_{\bf k}(t)}}\right|^{2} \nonumber \\
    &=\left[\dfrac{1}{2}\cos {\mu_3 t}+\dfrac{1}{N_1^{2}}\left[4\vert \alpha_{\bf k} \vert^{2} + {\beta^2} \cos {\mu_5 t}\right]\right]^{2} \label{ea-2p-qkt}.
\end{align}
Taking the long-time average of $q_{\bf k}(t)$, we get
\begin{align}
  \bar q_{\bf k} &= \lim_{T\rightarrow \infty} \frac{1}{T} \int_0^T dt~ q_{\bf k}(t), \nonumber\\
  &=\dfrac{1}{8}+\dfrac{1}{8}\left[\dfrac{32\vert \alpha_{\bf k} \vert^{4}+\beta^{4}}{(4\vert \alpha_{\bf k}\vert^{2}+\beta^{2})^2} \right].
  \label{ea-2p-qbar}
\end{align}

  
\bibliography{reference}

\begin{thebibliography}{57}%
\makeatletter
\providecommand \@ifxundefined [1]{%
 \@ifx{#1\undefined}
}%
\providecommand \@ifnum [1]{%
 \ifnum #1\expandafter \@firstoftwo
 \else \expandafter \@secondoftwo
 \fi
}%
\providecommand \@ifx [1]{%
 \ifx #1\expandafter \@firstoftwo
 \else \expandafter \@secondoftwo
 \fi
}%
\providecommand \natexlab [1]{#1}%
\providecommand \enquote  [1]{``#1''}%
\providecommand \bibnamefont  [1]{#1}%
\providecommand \bibfnamefont [1]{#1}%
\providecommand \citenamefont [1]{#1}%
\providecommand \href@noop [0]{\@secondoftwo}%
\providecommand \href [0]{\begingroup \@sanitize@url \@href}%
\providecommand \@href[1]{\@@startlink{#1}\@@href}%
\providecommand \@@href[1]{\endgroup#1\@@endlink}%
\providecommand \@sanitize@url [0]{\catcode `\\12\catcode `\$12\catcode
  `\&12\catcode `\#12\catcode `\^12\catcode `\_12\catcode `\%12\relax}%
\providecommand \@@startlink[1]{}%
\providecommand \@@endlink[0]{}%
\providecommand \url  [0]{\begingroup\@sanitize@url \@url }%
\providecommand \@url [1]{\endgroup\@href {#1}{\urlprefix }}%
\providecommand \urlprefix  [0]{URL }%
\providecommand \Eprint [0]{\href }%
\providecommand \doibase [0]{http://dx.doi.org/}%
\providecommand \selectlanguage [0]{\@gobble}%
\providecommand \bibinfo  [0]{\@secondoftwo}%
\providecommand \bibfield  [0]{\@secondoftwo}%
\providecommand \translation [1]{[#1]}%
\providecommand \BibitemOpen [0]{}%
\providecommand \bibitemStop [0]{}%
\providecommand \bibitemNoStop [0]{.\EOS\space}%
\providecommand \EOS [0]{\spacefactor3000\relax}%
\providecommand \BibitemShut  [1]{\csname bibitem#1\endcsname}%
\let\auto@bib@innerbib\@empty
\bibitem [{\citenamefont {Bastidas}\ \emph
  {et~al.}(2012{\natexlab{a}})\citenamefont {Bastidas}, \citenamefont {Emary},
  \citenamefont {Regler},\ and\ \citenamefont {Brandes}}]{BasEma(12a)}%
  \BibitemOpen
  \bibfield  {author} {\bibinfo {author} {\bibfnamefont {V.~M.}\ \bibnamefont
  {Bastidas}}, \bibinfo {author} {\bibfnamefont {C.}~\bibnamefont {Emary}},
  \bibinfo {author} {\bibfnamefont {B.}~\bibnamefont {Regler}}, \ and\ \bibinfo
  {author} {\bibfnamefont {T.}~\bibnamefont {Brandes}},\ }\href {\doibase
  10.1103/PhysRevLett.108.043003} {\bibfield  {journal} {\bibinfo  {journal}
  {Phys. Rev. Lett.}\ }\textbf {\bibinfo {volume} {108}},\ \bibinfo {pages}
  {043003} (\bibinfo {year} {2012}{\natexlab{a}})}\BibitemShut {NoStop}%
\bibitem [{\citenamefont {Bastidas}\ \emph
  {et~al.}(2012{\natexlab{b}})\citenamefont {Bastidas}, \citenamefont {Emary},
  \citenamefont {Schaller},\ and\ \citenamefont {Brandes}}]{BasEma(12b)}%
  \BibitemOpen
  \bibfield  {author} {\bibinfo {author} {\bibfnamefont {V.~M.}\ \bibnamefont
  {Bastidas}}, \bibinfo {author} {\bibfnamefont {C.}~\bibnamefont {Emary}},
  \bibinfo {author} {\bibfnamefont {G.}~\bibnamefont {Schaller}}, \ and\
  \bibinfo {author} {\bibfnamefont {T.}~\bibnamefont {Brandes}},\ }\href
  {\doibase 10.1103/PhysRevA.86.063627} {\bibfield  {journal} {\bibinfo
  {journal} {Phys. Rev. A}\ }\textbf {\bibinfo {volume} {86}},\ \bibinfo
  {pages} {063627} (\bibinfo {year} {2012}{\natexlab{b}})}\BibitemShut
  {NoStop}%
\bibitem [{\citenamefont {Bastidas}\ \emph {et~al.}(2014)\citenamefont
  {Bastidas}, \citenamefont {Perez-Fernandez}, \citenamefont {Vogl},\ and\
  \citenamefont {Brandes}}]{BasPer(14)}%
  \BibitemOpen
  \bibfield  {author} {\bibinfo {author} {\bibfnamefont {V.~M.}\ \bibnamefont
  {Bastidas}}, \bibinfo {author} {\bibfnamefont {P.}~\bibnamefont
  {Perez-Fernandez}}, \bibinfo {author} {\bibfnamefont {M.}~\bibnamefont
  {Vogl}}, \ and\ \bibinfo {author} {\bibfnamefont {T.}~\bibnamefont
  {Brandes}},\ }\href {\doibase 10.1103/physrevlett.112.140408} {\bibfield
  {journal} {\bibinfo  {journal} {Physical Review Letters}\ }\textbf {\bibinfo
  {volume} {112}} (\bibinfo {year} {2014}),\
  10.1103/physrevlett.112.140408}\BibitemShut {NoStop}%
\bibitem [{\citenamefont {Engelhardt}\ \emph {et~al.}(2013)\citenamefont
  {Engelhardt}, \citenamefont {Bastidas}, \citenamefont {Emary},\ and\
  \citenamefont {Brandes}}]{EngBas(13)}%
  \BibitemOpen
  \bibfield  {author} {\bibinfo {author} {\bibfnamefont {G.}~\bibnamefont
  {Engelhardt}}, \bibinfo {author} {\bibfnamefont {V.~M.}\ \bibnamefont
  {Bastidas}}, \bibinfo {author} {\bibfnamefont {C.}~\bibnamefont {Emary}}, \
  and\ \bibinfo {author} {\bibfnamefont {T.}~\bibnamefont {Brandes}},\ }\href
  {\doibase 10.1103/PhysRevE.87.052110} {\bibfield  {journal} {\bibinfo
  {journal} {Phys. Rev. E}\ }\textbf {\bibinfo {volume} {87}},\ \bibinfo
  {pages} {052110} (\bibinfo {year} {2013})}\BibitemShut {NoStop}%
\bibitem [{\citenamefont {Bukov}\ \emph {et~al.}(2015)\citenamefont {Bukov},
  \citenamefont {D'Alessio},\ and\ \citenamefont {Polkovnikov}}]{BukDal(15)}%
  \BibitemOpen
  \bibfield  {author} {\bibinfo {author} {\bibfnamefont {M.}~\bibnamefont
  {Bukov}}, \bibinfo {author} {\bibfnamefont {L.}~\bibnamefont {D'Alessio}}, \
  and\ \bibinfo {author} {\bibfnamefont {A.}~\bibnamefont {Polkovnikov}},\
  }\href {\doibase 10.1080/00018732.2015.1055918} {\bibfield  {journal}
  {\bibinfo  {journal} {Advances in Physics}\ }\textbf {\bibinfo {volume}
  {64}},\ \bibinfo {pages} {139} (\bibinfo {year} {2015})}\BibitemShut
  {NoStop}%
\bibitem [{\citenamefont {Goldman}\ and\ \citenamefont
  {Dalibard}(2014)}]{GolDal(14)}%
  \BibitemOpen
  \bibfield  {author} {\bibinfo {author} {\bibfnamefont {N.}~\bibnamefont
  {Goldman}}\ and\ \bibinfo {author} {\bibfnamefont {J.}~\bibnamefont
  {Dalibard}},\ }\href {\doibase 10.1103/PhysRevX.4.031027} {\bibfield
  {journal} {\bibinfo  {journal} {Phys. Rev. X}\ }\textbf {\bibinfo {volume}
  {4}},\ \bibinfo {pages} {031027} (\bibinfo {year} {2014})}\BibitemShut
  {NoStop}%
\bibitem [{\citenamefont {Das}(2010)}]{Das(10)}%
  \BibitemOpen
  \bibfield  {author} {\bibinfo {author} {\bibfnamefont {A.}~\bibnamefont
  {Das}},\ }\href {\doibase 10.1103/PhysRevB.82.172402} {\bibfield  {journal}
  {\bibinfo  {journal} {Phys. Rev. B}\ }\textbf {\bibinfo {volume} {82}},\
  \bibinfo {pages} {172402} (\bibinfo {year} {2010})}\BibitemShut {NoStop}%
\bibitem [{\citenamefont {Bhattacharyya}\ \emph {et~al.}(2012)\citenamefont
  {Bhattacharyya}, \citenamefont {Das},\ and\ \citenamefont
  {Dasgupta}}]{BhaDas(12)}%
  \BibitemOpen
  \bibfield  {author} {\bibinfo {author} {\bibfnamefont {S.}~\bibnamefont
  {Bhattacharyya}}, \bibinfo {author} {\bibfnamefont {A.}~\bibnamefont {Das}},
  \ and\ \bibinfo {author} {\bibfnamefont {S.}~\bibnamefont {Dasgupta}},\
  }\href {\doibase 10.1103/PhysRevB.86.054410} {\bibfield  {journal} {\bibinfo
  {journal} {Phys. Rev. B}\ }\textbf {\bibinfo {volume} {86}},\ \bibinfo
  {pages} {054410} (\bibinfo {year} {2012})}\BibitemShut {NoStop}%
\bibitem [{\citenamefont {Russomanno}\ \emph {et~al.}(2012)\citenamefont
  {Russomanno}, \citenamefont {Silva},\ and\ \citenamefont
  {Santoro}}]{RusSil(12)}%
  \BibitemOpen
  \bibfield  {author} {\bibinfo {author} {\bibfnamefont {A.}~\bibnamefont
  {Russomanno}}, \bibinfo {author} {\bibfnamefont {A.}~\bibnamefont {Silva}}, \
  and\ \bibinfo {author} {\bibfnamefont {G.~E.}\ \bibnamefont {Santoro}},\
  }\href {\doibase 10.1103/PhysRevLett.109.257201} {\bibfield  {journal}
  {\bibinfo  {journal} {Phys. Rev. Lett.}\ }\textbf {\bibinfo {volume} {109}},\
  \bibinfo {pages} {257201} (\bibinfo {year} {2012})}\BibitemShut {NoStop}%
\bibitem [{\citenamefont {Dunlap}\ and\ \citenamefont
  {Kenkre}(1986)}]{DunKen(86)}%
  \BibitemOpen
  \bibfield  {author} {\bibinfo {author} {\bibfnamefont {D.~H.}\ \bibnamefont
  {Dunlap}}\ and\ \bibinfo {author} {\bibfnamefont {V.~M.}\ \bibnamefont
  {Kenkre}},\ }\href {\doibase 10.1103/PhysRevB.34.3625} {\bibfield  {journal}
  {\bibinfo  {journal} {Phys. Rev. B}\ }\textbf {\bibinfo {volume} {34}},\
  \bibinfo {pages} {3625} (\bibinfo {year} {1986})}\BibitemShut {NoStop}%
\bibitem [{\citenamefont {Eckardt}\ \emph {et~al.}(2009)\citenamefont
  {Eckardt}, \citenamefont {Holthaus}, \citenamefont {Lignier}, \citenamefont
  {Zenesini}, \citenamefont {Ciampini}, \citenamefont {Morsch},\ and\
  \citenamefont {Arimondo}}]{EckHol(09)}%
  \BibitemOpen
  \bibfield  {author} {\bibinfo {author} {\bibfnamefont {A.}~\bibnamefont
  {Eckardt}}, \bibinfo {author} {\bibfnamefont {M.}~\bibnamefont {Holthaus}},
  \bibinfo {author} {\bibfnamefont {H.}~\bibnamefont {Lignier}}, \bibinfo
  {author} {\bibfnamefont {A.}~\bibnamefont {Zenesini}}, \bibinfo {author}
  {\bibfnamefont {D.}~\bibnamefont {Ciampini}}, \bibinfo {author}
  {\bibfnamefont {O.}~\bibnamefont {Morsch}}, \ and\ \bibinfo {author}
  {\bibfnamefont {E.}~\bibnamefont {Arimondo}},\ }\href {\doibase
  10.1103/PhysRevA.79.013611} {\bibfield  {journal} {\bibinfo  {journal} {Phys.
  Rev. A}\ }\textbf {\bibinfo {volume} {79}},\ \bibinfo {pages} {013611}
  (\bibinfo {year} {2009})}\BibitemShut {NoStop}%
\bibitem [{\citenamefont {Grossmann}\ \emph {et~al.}(1991)\citenamefont
  {Grossmann}, \citenamefont {Dittrich}, \citenamefont {Jung},\ and\
  \citenamefont {H\"anggi}}]{GroDit(91)}%
  \BibitemOpen
  \bibfield  {author} {\bibinfo {author} {\bibfnamefont {F.}~\bibnamefont
  {Grossmann}}, \bibinfo {author} {\bibfnamefont {T.}~\bibnamefont {Dittrich}},
  \bibinfo {author} {\bibfnamefont {P.}~\bibnamefont {Jung}}, \ and\ \bibinfo
  {author} {\bibfnamefont {P.}~\bibnamefont {H\"anggi}},\ }\href {\doibase
  10.1103/PhysRevLett.67.516} {\bibfield  {journal} {\bibinfo  {journal} {Phys.
  Rev. Lett.}\ }\textbf {\bibinfo {volume} {67}},\ \bibinfo {pages} {516}
  (\bibinfo {year} {1991})}\BibitemShut {NoStop}%
\bibitem [{\citenamefont {Gro{\ss}mann}\ and\ \citenamefont
  {HÃ€nggi}(1992)}]{GroHan(92)}%
  \BibitemOpen
  \bibfield  {author} {\bibinfo {author} {\bibfnamefont {F.}~\bibnamefont
  {Gro{\ss}mann}}\ and\ \bibinfo {author} {\bibfnamefont {P.}~\bibnamefont
  {HÃ€nggi}},\ }\href {\doibase 10.1209/0295-5075/18/7/001} {\bibfield
  {journal} {\bibinfo  {journal} {Europhysics Letters ({EPL})}\ }\textbf
  {\bibinfo {volume} {18}},\ \bibinfo {pages} {571} (\bibinfo {year}
  {1992})}\BibitemShut {NoStop}%
\bibitem [{\citenamefont {Roy}\ and\ \citenamefont {Das}(2015)}]{RoyDas(15)}%
  \BibitemOpen
  \bibfield  {author} {\bibinfo {author} {\bibfnamefont {A.}~\bibnamefont
  {Roy}}\ and\ \bibinfo {author} {\bibfnamefont {A.}~\bibnamefont {Das}},\
  }\href {\doibase 10.1103/PhysRevB.91.121106} {\bibfield  {journal} {\bibinfo
  {journal} {Phys. Rev. B}\ }\textbf {\bibinfo {volume} {91}},\ \bibinfo
  {pages} {121106} (\bibinfo {year} {2015})}\BibitemShut {NoStop}%
\bibitem [{\citenamefont {Haldar}\ \emph {et~al.}(2021)\citenamefont {Haldar},
  \citenamefont {Sen}, \citenamefont {Moessner},\ and\ \citenamefont
  {Das}}]{HalSen(21)}%
  \BibitemOpen
  \bibfield  {author} {\bibinfo {author} {\bibfnamefont {A.}~\bibnamefont
  {Haldar}}, \bibinfo {author} {\bibfnamefont {D.}~\bibnamefont {Sen}},
  \bibinfo {author} {\bibfnamefont {R.}~\bibnamefont {Moessner}}, \ and\
  \bibinfo {author} {\bibfnamefont {A.}~\bibnamefont {Das}},\ }\href {\doibase
  10.1103/PhysRevX.11.021008} {\bibfield  {journal} {\bibinfo  {journal} {Phys.
  Rev. X}\ }\textbf {\bibinfo {volume} {11}},\ \bibinfo {pages} {021008}
  (\bibinfo {year} {2021})}\BibitemShut {NoStop}%
\bibitem [{\citenamefont {{Das}}\ and\ \citenamefont
  {{Moessner}}()}]{DasMoe(12)}%
  \BibitemOpen
  \bibfield  {author} {\bibinfo {author} {\bibfnamefont {A.}~\bibnamefont
  {{Das}}}\ and\ \bibinfo {author} {\bibfnamefont {R.}~\bibnamefont
  {{Moessner}}},\ }\href@noop {} {}\Eprint {http://arxiv.org/abs/1208.0217}
  {arXiv:1208.0217} \BibitemShut {NoStop}%
\bibitem [{\citenamefont {Hegde}\ \emph {et~al.}(2014)\citenamefont {Hegde},
  \citenamefont {Katiyar}, \citenamefont {Mahesh},\ and\ \citenamefont
  {Das}}]{HegKat(14)}%
  \BibitemOpen
  \bibfield  {author} {\bibinfo {author} {\bibfnamefont {S.~S.}\ \bibnamefont
  {Hegde}}, \bibinfo {author} {\bibfnamefont {H.}~\bibnamefont {Katiyar}},
  \bibinfo {author} {\bibfnamefont {T.~S.}\ \bibnamefont {Mahesh}}, \ and\
  \bibinfo {author} {\bibfnamefont {A.}~\bibnamefont {Das}},\ }\href {\doibase
  10.1103/PhysRevB.90.174407} {\bibfield  {journal} {\bibinfo  {journal} {Phys.
  Rev. B}\ }\textbf {\bibinfo {volume} {90}},\ \bibinfo {pages} {174407}
  (\bibinfo {year} {2014})}\BibitemShut {NoStop}%
\bibitem [{\citenamefont {Haldar}\ and\ \citenamefont
  {Das}(2017)}]{HalDas(17)}%
  \BibitemOpen
  \bibfield  {author} {\bibinfo {author} {\bibfnamefont {A.}~\bibnamefont
  {Haldar}}\ and\ \bibinfo {author} {\bibfnamefont {A.}~\bibnamefont {Das}},\
  }\href {\doibase 10.1002/andp.201600333} {\bibfield  {journal} {\bibinfo
  {journal} {Annalen der Physik}\ }\textbf {\bibinfo {volume} {529}},\ \bibinfo
  {pages} {1600333} (\bibinfo {year} {2017})}\BibitemShut {NoStop}%
\bibitem [{\citenamefont {Haldar}\ and\ \citenamefont
  {Das}(2021)}]{HalDas(21)}%
  \BibitemOpen
  \bibfield  {author} {\bibinfo {author} {\bibfnamefont {A.}~\bibnamefont
  {Haldar}}\ and\ \bibinfo {author} {\bibfnamefont {A.}~\bibnamefont {Das}},\
  }\href@noop {} {\bibfield  {journal} {\bibinfo  {journal} {Journal of
  physics. Condensed matter : an Institute of Physics journal}\ } (\bibinfo
  {year} {2021})}\BibitemShut {NoStop}%
\bibitem [{\citenamefont {Abergel}\ \emph {et~al.}(2010)\citenamefont
  {Abergel}, \citenamefont {Apalkov}, \citenamefont {Berashevich},
  \citenamefont {Ziegler},\ and\ \citenamefont {Chakraborty}}]{AbeApa(10)}%
  \BibitemOpen
  \bibfield  {author} {\bibinfo {author} {\bibfnamefont {D.}~\bibnamefont
  {Abergel}}, \bibinfo {author} {\bibfnamefont {V.}~\bibnamefont {Apalkov}},
  \bibinfo {author} {\bibfnamefont {J.}~\bibnamefont {Berashevich}}, \bibinfo
  {author} {\bibfnamefont {K.}~\bibnamefont {Ziegler}}, \ and\ \bibinfo
  {author} {\bibfnamefont {T.}~\bibnamefont {Chakraborty}},\ }\href {\doibase
  10.1080/00018732.2010.487978} {\bibfield  {journal} {\bibinfo  {journal}
  {Advances in Physics}\ }\textbf {\bibinfo {volume} {59}},\ \bibinfo {pages}
  {261} (\bibinfo {year} {2010})},\ \Eprint
  {http://arxiv.org/abs/https://doi.org/10.1080/00018732.2010.487978}
  {https://doi.org/10.1080/00018732.2010.487978} \BibitemShut {NoStop}%
\bibitem [{\citenamefont {Konschuh}\ \emph {et~al.}(2012)\citenamefont
  {Konschuh}, \citenamefont {Gmitra}, \citenamefont {Kochan},\ and\
  \citenamefont {Fabian}}]{KonGmi(12)}%
  \BibitemOpen
  \bibfield  {author} {\bibinfo {author} {\bibfnamefont {S.}~\bibnamefont
  {Konschuh}}, \bibinfo {author} {\bibfnamefont {M.}~\bibnamefont {Gmitra}},
  \bibinfo {author} {\bibfnamefont {D.}~\bibnamefont {Kochan}}, \ and\ \bibinfo
  {author} {\bibfnamefont {J.}~\bibnamefont {Fabian}},\ }\href {\doibase
  10.1103/PhysRevB.85.115423} {\bibfield  {journal} {\bibinfo  {journal} {Phys.
  Rev. B}\ }\textbf {\bibinfo {volume} {85}},\ \bibinfo {pages} {115423}
  (\bibinfo {year} {2012})}\BibitemShut {NoStop}%
\bibitem [{\citenamefont {Oostinga}\ \emph {et~al.}(2008)\citenamefont
  {Oostinga}, \citenamefont {Heersche}, \citenamefont {Liu}, \citenamefont
  {Morpurgo},\ and\ \citenamefont {Vandersypen}}]{OosHee(08)}%
  \BibitemOpen
  \bibfield  {author} {\bibinfo {author} {\bibfnamefont {J.~B.}\ \bibnamefont
  {Oostinga}}, \bibinfo {author} {\bibfnamefont {H.~B.}\ \bibnamefont
  {Heersche}}, \bibinfo {author} {\bibfnamefont {X.}~\bibnamefont {Liu}},
  \bibinfo {author} {\bibfnamefont {A.~F.}\ \bibnamefont {Morpurgo}}, \ and\
  \bibinfo {author} {\bibfnamefont {L.~M.~K.}\ \bibnamefont {Vandersypen}},\
  }\href {\doibase 10.1038/nmat2082} {\bibfield  {journal} {\bibinfo  {journal}
  {Nature Materials}\ }\textbf {\bibinfo {volume} {7}},\ \bibinfo {pages} {151}
  (\bibinfo {year} {2008})}\BibitemShut {NoStop}%
\bibitem [{\citenamefont {Gosciniak}\ and\ \citenamefont
  {Tan}(2013)}]{GosTan(13)}%
  \BibitemOpen
  \bibfield  {author} {\bibinfo {author} {\bibfnamefont {J.}~\bibnamefont
  {Gosciniak}}\ and\ \bibinfo {author} {\bibfnamefont {D.~T.~H.}\ \bibnamefont
  {Tan}},\ }\href {\doibase 10.1038/srep01897} {\bibfield  {journal} {\bibinfo
  {journal} {Scientific Reports}\ }\textbf {\bibinfo {volume} {3}},\ \bibinfo
  {pages} {1897} (\bibinfo {year} {2013})}\BibitemShut {NoStop}%
\bibitem [{\citenamefont {Zhang}\ \emph {et~al.}(2011)\citenamefont {Zhang},
  \citenamefont {Lin}, \citenamefont {Liu}, \citenamefont {Tite}, \citenamefont
  {Su}, \citenamefont {Chang}, \citenamefont {Lee}, \citenamefont {Chu},
  \citenamefont {Wei}, \citenamefont {Kuo},\ and\ \citenamefont
  {Li}}]{ZhaLin(11)}%
  \BibitemOpen
  \bibfield  {author} {\bibinfo {author} {\bibfnamefont {W.}~\bibnamefont
  {Zhang}}, \bibinfo {author} {\bibfnamefont {C.-T.}\ \bibnamefont {Lin}},
  \bibinfo {author} {\bibfnamefont {K.-K.}\ \bibnamefont {Liu}}, \bibinfo
  {author} {\bibfnamefont {T.}~\bibnamefont {Tite}}, \bibinfo {author}
  {\bibfnamefont {C.-Y.}\ \bibnamefont {Su}}, \bibinfo {author} {\bibfnamefont
  {C.-H.}\ \bibnamefont {Chang}}, \bibinfo {author} {\bibfnamefont {Y.-H.}\
  \bibnamefont {Lee}}, \bibinfo {author} {\bibfnamefont {C.-W.}\ \bibnamefont
  {Chu}}, \bibinfo {author} {\bibfnamefont {K.-H.}\ \bibnamefont {Wei}},
  \bibinfo {author} {\bibfnamefont {J.-L.}\ \bibnamefont {Kuo}}, \ and\
  \bibinfo {author} {\bibfnamefont {L.-J.}\ \bibnamefont {Li}},\ }\href
  {\doibase 10.1021/nn202463g} {\bibfield  {journal} {\bibinfo  {journal} {ACS
  Nano}\ }\textbf {\bibinfo {volume} {5}},\ \bibinfo {pages} {7517} (\bibinfo
  {year} {2011})}\BibitemShut {NoStop}%
\bibitem [{\citenamefont {Bostwick}\ \emph {et~al.}(2009)\citenamefont
  {Bostwick}, \citenamefont {McChesney}, \citenamefont {Ohta}, \citenamefont
  {Rotenberg}, \citenamefont {Seyller},\ and\ \citenamefont
  {Horn}}]{BosMcc(09)}%
  \BibitemOpen
  \bibfield  {author} {\bibinfo {author} {\bibfnamefont {A.}~\bibnamefont
  {Bostwick}}, \bibinfo {author} {\bibfnamefont {J.}~\bibnamefont {McChesney}},
  \bibinfo {author} {\bibfnamefont {T.}~\bibnamefont {Ohta}}, \bibinfo {author}
  {\bibfnamefont {E.}~\bibnamefont {Rotenberg}}, \bibinfo {author}
  {\bibfnamefont {T.}~\bibnamefont {Seyller}}, \ and\ \bibinfo {author}
  {\bibfnamefont {K.}~\bibnamefont {Horn}},\ }\href {\doibase
  https://doi.org/10.1016/j.progsurf.2009.08.002} {\bibfield  {journal}
  {\bibinfo  {journal} {Progress in Surface Science}\ }\textbf {\bibinfo
  {volume} {84}},\ \bibinfo {pages} {380} (\bibinfo {year} {2009})}\BibitemShut
  {NoStop}%
\bibitem [{\citenamefont {Zhang}\ \emph {et~al.}(2008)\citenamefont {Zhang},
  \citenamefont {Li}, \citenamefont {Basov}, \citenamefont {Fogler},
  \citenamefont {Hao},\ and\ \citenamefont {Martin}}]{ZhaBas(08)}%
  \BibitemOpen
  \bibfield  {author} {\bibinfo {author} {\bibfnamefont {L.~M.}\ \bibnamefont
  {Zhang}}, \bibinfo {author} {\bibfnamefont {Z.~Q.}\ \bibnamefont {Li}},
  \bibinfo {author} {\bibfnamefont {D.~N.}\ \bibnamefont {Basov}}, \bibinfo
  {author} {\bibfnamefont {M.~M.}\ \bibnamefont {Fogler}}, \bibinfo {author}
  {\bibfnamefont {Z.}~\bibnamefont {Hao}}, \ and\ \bibinfo {author}
  {\bibfnamefont {M.~C.}\ \bibnamefont {Martin}},\ }\href {\doibase
  10.1103/PhysRevB.78.235408} {\bibfield  {journal} {\bibinfo  {journal} {Phys.
  Rev. B}\ }\textbf {\bibinfo {volume} {78}},\ \bibinfo {pages} {235408}
  (\bibinfo {year} {2008})}\BibitemShut {NoStop}%
\bibitem [{\citenamefont {Feldman}\ \emph {et~al.}(2009)\citenamefont
  {Feldman}, \citenamefont {Martin},\ and\ \citenamefont
  {Yacoby}}]{FelMar(09)}%
  \BibitemOpen
  \bibfield  {author} {\bibinfo {author} {\bibfnamefont {B.~E.}\ \bibnamefont
  {Feldman}}, \bibinfo {author} {\bibfnamefont {J.}~\bibnamefont {Martin}}, \
  and\ \bibinfo {author} {\bibfnamefont {A.}~\bibnamefont {Yacoby}},\ }\href
  {\doibase 10.1038/nphys1406} {\bibfield  {journal} {\bibinfo  {journal}
  {Nature Physics}\ }\textbf {\bibinfo {volume} {5}},\ \bibinfo {pages} {889}
  (\bibinfo {year} {2009})}\BibitemShut {NoStop}%
\bibitem [{\citenamefont {Min}\ \emph {et~al.}(2007)\citenamefont {Min},
  \citenamefont {Sahu}, \citenamefont {Banerjee},\ and\ \citenamefont
  {MacDonald}}]{MinSah(07)}%
  \BibitemOpen
  \bibfield  {author} {\bibinfo {author} {\bibfnamefont {H.}~\bibnamefont
  {Min}}, \bibinfo {author} {\bibfnamefont {B.}~\bibnamefont {Sahu}}, \bibinfo
  {author} {\bibfnamefont {S.~K.}\ \bibnamefont {Banerjee}}, \ and\ \bibinfo
  {author} {\bibfnamefont {A.~H.}\ \bibnamefont {MacDonald}},\ }\href {\doibase
  10.1103/PhysRevB.75.155115} {\bibfield  {journal} {\bibinfo  {journal} {Phys.
  Rev. B}\ }\textbf {\bibinfo {volume} {75}},\ \bibinfo {pages} {155115}
  (\bibinfo {year} {2007})}\BibitemShut {NoStop}%
\bibitem [{\citenamefont {McCann}(2006)}]{MccEdw(06)}%
  \BibitemOpen
  \bibfield  {author} {\bibinfo {author} {\bibfnamefont {E.}~\bibnamefont
  {McCann}},\ }\href {\doibase 10.1103/PhysRevB.74.161403} {\bibfield
  {journal} {\bibinfo  {journal} {Phys. Rev. B}\ }\textbf {\bibinfo {volume}
  {74}},\ \bibinfo {pages} {161403} (\bibinfo {year} {2006})}\BibitemShut
  {NoStop}%
\bibitem [{\citenamefont {Lu}\ \emph {et~al.}(2006)\citenamefont {Lu},
  \citenamefont {Chang}, \citenamefont {Huang}, \citenamefont {Chen},\ and\
  \citenamefont {Lin}}]{ChaHua(06)}%
  \BibitemOpen
  \bibfield  {author} {\bibinfo {author} {\bibfnamefont {C.~L.}\ \bibnamefont
  {Lu}}, \bibinfo {author} {\bibfnamefont {C.~P.}\ \bibnamefont {Chang}},
  \bibinfo {author} {\bibfnamefont {Y.~C.}\ \bibnamefont {Huang}}, \bibinfo
  {author} {\bibfnamefont {R.~B.}\ \bibnamefont {Chen}}, \ and\ \bibinfo
  {author} {\bibfnamefont {M.~L.}\ \bibnamefont {Lin}},\ }\href {\doibase
  10.1103/PhysRevB.73.144427} {\bibfield  {journal} {\bibinfo  {journal} {Phys.
  Rev. B}\ }\textbf {\bibinfo {volume} {73}},\ \bibinfo {pages} {144427}
  (\bibinfo {year} {2006})}\BibitemShut {NoStop}%
\bibitem [{\citenamefont {Zhang}\ \emph {et~al.}(2009)\citenamefont {Zhang},
  \citenamefont {Tang}, \citenamefont {Girit}, \citenamefont {Hao},
  \citenamefont {Martin}, \citenamefont {Zettl}, \citenamefont {Crommie},
  \citenamefont {Shen},\ and\ \citenamefont {Wang}}]{ZhaYua(09)}%
  \BibitemOpen
  \bibfield  {author} {\bibinfo {author} {\bibfnamefont {Y.}~\bibnamefont
  {Zhang}}, \bibinfo {author} {\bibfnamefont {T.-T.}\ \bibnamefont {Tang}},
  \bibinfo {author} {\bibfnamefont {C.}~\bibnamefont {Girit}}, \bibinfo
  {author} {\bibfnamefont {Z.}~\bibnamefont {Hao}}, \bibinfo {author}
  {\bibfnamefont {M.~C.}\ \bibnamefont {Martin}}, \bibinfo {author}
  {\bibfnamefont {A.}~\bibnamefont {Zettl}}, \bibinfo {author} {\bibfnamefont
  {M.~F.}\ \bibnamefont {Crommie}}, \bibinfo {author} {\bibfnamefont {Y.~R.}\
  \bibnamefont {Shen}}, \ and\ \bibinfo {author} {\bibfnamefont
  {F.}~\bibnamefont {Wang}},\ }\href {\doibase 10.1038/nature08105} {\bibfield
  {journal} {\bibinfo  {journal} {Nature}\ }\textbf {\bibinfo {volume} {459}},\
  \bibinfo {pages} {820} (\bibinfo {year} {2009})}\BibitemShut {NoStop}%
\bibitem [{\citenamefont {Castro}\ \emph {et~al.}(2007)\citenamefont {Castro},
  \citenamefont {Novoselov}, \citenamefont {Morozov}, \citenamefont {Peres},
  \citenamefont {dos Santos}, \citenamefont {Nilsson}, \citenamefont {Guinea},
  \citenamefont {Geim},\ and\ \citenamefont {Neto}}]{CasNov(07)}%
  \BibitemOpen
  \bibfield  {author} {\bibinfo {author} {\bibfnamefont {E.~V.}\ \bibnamefont
  {Castro}}, \bibinfo {author} {\bibfnamefont {K.~S.}\ \bibnamefont
  {Novoselov}}, \bibinfo {author} {\bibfnamefont {S.~V.}\ \bibnamefont
  {Morozov}}, \bibinfo {author} {\bibfnamefont {N.~M.~R.}\ \bibnamefont
  {Peres}}, \bibinfo {author} {\bibfnamefont {J.~M. B.~L.}\ \bibnamefont {dos
  Santos}}, \bibinfo {author} {\bibfnamefont {J.}~\bibnamefont {Nilsson}},
  \bibinfo {author} {\bibfnamefont {F.}~\bibnamefont {Guinea}}, \bibinfo
  {author} {\bibfnamefont {A.~K.}\ \bibnamefont {Geim}}, \ and\ \bibinfo
  {author} {\bibfnamefont {A.~H.~C.}\ \bibnamefont {Neto}},\ }\href {\doibase
  10.1103/PhysRevLett.99.216802} {\bibfield  {journal} {\bibinfo  {journal}
  {Phys. Rev. Lett.}\ }\textbf {\bibinfo {volume} {99}},\ \bibinfo {pages}
  {216802} (\bibinfo {year} {2007})}\BibitemShut {NoStop}%
\bibitem [{\citenamefont {Haldar}\ \emph {et~al.}(2018)\citenamefont {Haldar},
  \citenamefont {Moessner},\ and\ \citenamefont {Das}}]{HalMoe(18)}%
  \BibitemOpen
  \bibfield  {author} {\bibinfo {author} {\bibfnamefont {A.}~\bibnamefont
  {Haldar}}, \bibinfo {author} {\bibfnamefont {R.}~\bibnamefont {Moessner}}, \
  and\ \bibinfo {author} {\bibfnamefont {A.}~\bibnamefont {Das}},\ }\href
  {\doibase 10.1103/PhysRevB.97.245122} {\bibfield  {journal} {\bibinfo
  {journal} {Phys. Rev. B}\ }\textbf {\bibinfo {volume} {97}},\ \bibinfo
  {pages} {245122} (\bibinfo {year} {2018})}\BibitemShut {NoStop}%
\bibitem [{\citenamefont {Lignier}\ \emph {et~al.}(2007)\citenamefont
  {Lignier}, \citenamefont {Sias}, \citenamefont {Ciampini}, \citenamefont
  {Singh}, \citenamefont {Zenesini}, \citenamefont {Morsch},\ and\
  \citenamefont {Arimondo}}]{LigSia(07)}%
  \BibitemOpen
  \bibfield  {author} {\bibinfo {author} {\bibfnamefont {H.}~\bibnamefont
  {Lignier}}, \bibinfo {author} {\bibfnamefont {C.}~\bibnamefont {Sias}},
  \bibinfo {author} {\bibfnamefont {D.}~\bibnamefont {Ciampini}}, \bibinfo
  {author} {\bibfnamefont {Y.}~\bibnamefont {Singh}}, \bibinfo {author}
  {\bibfnamefont {A.}~\bibnamefont {Zenesini}}, \bibinfo {author}
  {\bibfnamefont {O.}~\bibnamefont {Morsch}}, \ and\ \bibinfo {author}
  {\bibfnamefont {E.}~\bibnamefont {Arimondo}},\ }\href {\doibase
  10.1103/PhysRevLett.99.220403} {\bibfield  {journal} {\bibinfo  {journal}
  {Phys. Rev. Lett.}\ }\textbf {\bibinfo {volume} {99}},\ \bibinfo {pages}
  {220403} (\bibinfo {year} {2007})}\BibitemShut {NoStop}%
\bibitem [{\citenamefont {Geim}\ and\ \citenamefont
  {Novoselov}(2007)}]{GeiNov(07)}%
  \BibitemOpen
  \bibfield  {author} {\bibinfo {author} {\bibfnamefont {A.~K.}\ \bibnamefont
  {Geim}}\ and\ \bibinfo {author} {\bibfnamefont {K.~S.}\ \bibnamefont
  {Novoselov}},\ }\href {\doibase 10.1038/nmat1849} {\bibfield  {journal}
  {\bibinfo  {journal} {Nature Materials}\ }\textbf {\bibinfo {volume} {6}},\
  \bibinfo {pages} {183} (\bibinfo {year} {2007})}\BibitemShut {NoStop}%
\bibitem [{\citenamefont {Novoselov}\ \emph {et~al.}(2004)\citenamefont
  {Novoselov}, \citenamefont {Geim}, \citenamefont {Morozov}, \citenamefont
  {Jiang}, \citenamefont {Zhang}, \citenamefont {Dubonos}, \citenamefont
  {Grigorieva},\ and\ \citenamefont {Firsov}}]{NovGei(04)}%
  \BibitemOpen
  \bibfield  {author} {\bibinfo {author} {\bibfnamefont {K.~S.}\ \bibnamefont
  {Novoselov}}, \bibinfo {author} {\bibfnamefont {A.~K.}\ \bibnamefont {Geim}},
  \bibinfo {author} {\bibfnamefont {S.~V.}\ \bibnamefont {Morozov}}, \bibinfo
  {author} {\bibfnamefont {D.}~\bibnamefont {Jiang}}, \bibinfo {author}
  {\bibfnamefont {Y.}~\bibnamefont {Zhang}}, \bibinfo {author} {\bibfnamefont
  {S.~V.}\ \bibnamefont {Dubonos}}, \bibinfo {author} {\bibfnamefont {I.~V.}\
  \bibnamefont {Grigorieva}}, \ and\ \bibinfo {author} {\bibfnamefont {A.~A.}\
  \bibnamefont {Firsov}},\ }\href {\doibase 10.1126/science.1102896} {\bibfield
   {journal} {\bibinfo  {journal} {Science}\ }\textbf {\bibinfo {volume}
  {306}},\ \bibinfo {pages} {666} (\bibinfo {year} {2004})}\BibitemShut
  {NoStop}%
\bibitem [{\citenamefont {Wang}\ \emph {et~al.}(2012)\citenamefont {Wang},
  \citenamefont {Guo}, \citenamefont {Liu},\ and\ \citenamefont
  {Sheng}}]{WanGuo(12)}%
  \BibitemOpen
  \bibfield  {author} {\bibinfo {author} {\bibfnamefont {T.}~\bibnamefont
  {Wang}}, \bibinfo {author} {\bibfnamefont {Q.}~\bibnamefont {Guo}}, \bibinfo
  {author} {\bibfnamefont {Y.}~\bibnamefont {Liu}}, \ and\ \bibinfo {author}
  {\bibfnamefont {K.}~\bibnamefont {Sheng}},\ }\href {\doibase
  10.1088/1674-1056/21/6/067301} {\bibfield  {journal} {\bibinfo  {journal}
  {Chinese Physics B}\ }\textbf {\bibinfo {volume} {21}},\ \bibinfo {pages}
  {067301} (\bibinfo {year} {2012})}\BibitemShut {NoStop}%
\bibitem [{\citenamefont {Rozhkov}\ \emph {et~al.}(2016)\citenamefont
  {Rozhkov}, \citenamefont {Sboychakov}, \citenamefont {Rakhmanov},\ and\
  \citenamefont {Nori}}]{RozSbo(16)}%
  \BibitemOpen
  \bibfield  {author} {\bibinfo {author} {\bibfnamefont {A.}~\bibnamefont
  {Rozhkov}}, \bibinfo {author} {\bibfnamefont {A.}~\bibnamefont {Sboychakov}},
  \bibinfo {author} {\bibfnamefont {A.}~\bibnamefont {Rakhmanov}}, \ and\
  \bibinfo {author} {\bibfnamefont {F.}~\bibnamefont {Nori}},\ }\href {\doibase
  10.1016/j.physrep.2016.07.003} {\bibfield  {journal} {\bibinfo  {journal}
  {Physics Reports}\ }\textbf {\bibinfo {volume} {648}},\ \bibinfo {pages} {1}
  (\bibinfo {year} {2016})}\BibitemShut {NoStop}%
\bibitem [{\citenamefont {Andrei}\ and\ \citenamefont
  {MacDonald}(2020)}]{AndMac(20)}%
  \BibitemOpen
  \bibfield  {author} {\bibinfo {author} {\bibfnamefont {E.~Y.}\ \bibnamefont
  {Andrei}}\ and\ \bibinfo {author} {\bibfnamefont {A.~H.}\ \bibnamefont
  {MacDonald}},\ }\href {\doibase 10.1038/s41563-020-00840-0} {\bibfield
  {journal} {\bibinfo  {journal} {Nature Materials}\ }\textbf {\bibinfo
  {volume} {19}},\ \bibinfo {pages} {1265} (\bibinfo {year}
  {2020})}\BibitemShut {NoStop}%
\bibitem [{\citenamefont {Choi}\ \emph {et~al.}(2019)\citenamefont {Choi},
  \citenamefont {Kemmer}, \citenamefont {Peng}, \citenamefont {Thomson},
  \citenamefont {Arora}, \citenamefont {Polski}, \citenamefont {Zhang},
  \citenamefont {Ren}, \citenamefont {Alicea}, \citenamefont {Refael},
  \citenamefont {von Oppen}, \citenamefont {Watanabe}, \citenamefont
  {Taniguchi},\ and\ \citenamefont {Nadj-Perge}}]{ChoKem(19)}%
  \BibitemOpen
  \bibfield  {author} {\bibinfo {author} {\bibfnamefont {Y.}~\bibnamefont
  {Choi}}, \bibinfo {author} {\bibfnamefont {J.}~\bibnamefont {Kemmer}},
  \bibinfo {author} {\bibfnamefont {Y.}~\bibnamefont {Peng}}, \bibinfo {author}
  {\bibfnamefont {A.}~\bibnamefont {Thomson}}, \bibinfo {author} {\bibfnamefont
  {H.}~\bibnamefont {Arora}}, \bibinfo {author} {\bibfnamefont
  {R.}~\bibnamefont {Polski}}, \bibinfo {author} {\bibfnamefont
  {Y.}~\bibnamefont {Zhang}}, \bibinfo {author} {\bibfnamefont
  {H.}~\bibnamefont {Ren}}, \bibinfo {author} {\bibfnamefont {J.}~\bibnamefont
  {Alicea}}, \bibinfo {author} {\bibfnamefont {G.}~\bibnamefont {Refael}},
  \bibinfo {author} {\bibfnamefont {F.}~\bibnamefont {von Oppen}}, \bibinfo
  {author} {\bibfnamefont {K.}~\bibnamefont {Watanabe}}, \bibinfo {author}
  {\bibfnamefont {T.}~\bibnamefont {Taniguchi}}, \ and\ \bibinfo {author}
  {\bibfnamefont {S.}~\bibnamefont {Nadj-Perge}},\ }\href {\doibase
  10.1038/s41567-019-0606-5} {\bibfield  {journal} {\bibinfo  {journal} {Nature
  Physics}\ }\textbf {\bibinfo {volume} {15}},\ \bibinfo {pages} {1174}
  (\bibinfo {year} {2019})}\BibitemShut {NoStop}%
\bibitem [{\citenamefont {McCann}\ \emph {et~al.}(2007)\citenamefont {McCann},
  \citenamefont {Abergel},\ and\ \citenamefont {Fal'ko}}]{McCAbe(07)}%
  \BibitemOpen
  \bibfield  {author} {\bibinfo {author} {\bibfnamefont {E.}~\bibnamefont
  {McCann}}, \bibinfo {author} {\bibfnamefont {D.~S.~L.}\ \bibnamefont
  {Abergel}}, \ and\ \bibinfo {author} {\bibfnamefont {V.~I.}\ \bibnamefont
  {Fal'ko}},\ }\href {\doibase 10.1140/epjst/e2007-00229-1} {\bibfield
  {journal} {\bibinfo  {journal} {The European Physical Journal Special
  Topics}\ }\textbf {\bibinfo {volume} {148}},\ \bibinfo {pages} {91} (\bibinfo
  {year} {2007})}\BibitemShut {NoStop}%
\bibitem [{\citenamefont {Moon}\ and\ \citenamefont
  {Koshino}(2012)}]{MooKosh(12)}%
  \BibitemOpen
  \bibfield  {author} {\bibinfo {author} {\bibfnamefont {P.}~\bibnamefont
  {Moon}}\ and\ \bibinfo {author} {\bibfnamefont {M.}~\bibnamefont {Koshino}},\
  }\href {\doibase 10.1103/PhysRevB.85.195458} {\bibfield  {journal} {\bibinfo
  {journal} {Phys. Rev. B}\ }\textbf {\bibinfo {volume} {85}},\ \bibinfo
  {pages} {195458} (\bibinfo {year} {2012})}\BibitemShut {NoStop}%
\bibitem [{\citenamefont {Yan}\ \emph {et~al.}(2011)\citenamefont {Yan},
  \citenamefont {Peng}, \citenamefont {Zhou}, \citenamefont {Li},\ and\
  \citenamefont {Liu}}]{YanPen(11)}%
  \BibitemOpen
  \bibfield  {author} {\bibinfo {author} {\bibfnamefont {K.}~\bibnamefont
  {Yan}}, \bibinfo {author} {\bibfnamefont {H.}~\bibnamefont {Peng}}, \bibinfo
  {author} {\bibfnamefont {Y.}~\bibnamefont {Zhou}}, \bibinfo {author}
  {\bibfnamefont {H.}~\bibnamefont {Li}}, \ and\ \bibinfo {author}
  {\bibfnamefont {Z.}~\bibnamefont {Liu}},\ }\href {\doibase 10.1021/nl104000b}
  {\bibfield  {journal} {\bibinfo  {journal} {Nano Letters}\ }\textbf {\bibinfo
  {volume} {11}},\ \bibinfo {pages} {1106} (\bibinfo {year} {2011})},\ \bibinfo
  {note} {pMID: 21322597},\ \Eprint
  {http://arxiv.org/abs/https://doi.org/10.1021/nl104000b}
  {https://doi.org/10.1021/nl104000b} \BibitemShut {NoStop}%
\bibitem [{\citenamefont {Ould~NE}\ \emph {et~al.}(2017)\citenamefont
  {Ould~NE}, \citenamefont {Boujnah}, \citenamefont {Benyoussef},\ and\
  \citenamefont {Kenz}}]{OulBou(17)}%
  \BibitemOpen
  \bibfield  {author} {\bibinfo {author} {\bibfnamefont {M.~L.}\ \bibnamefont
  {Ould~NE}}, \bibinfo {author} {\bibfnamefont {M.}~\bibnamefont {Boujnah}},
  \bibinfo {author} {\bibfnamefont {A.}~\bibnamefont {Benyoussef}}, \ and\
  \bibinfo {author} {\bibfnamefont {A.~E.}\ \bibnamefont {Kenz}},\ }\href
  {\doibase 10.1007/s10948-016-3910-7} {\bibfield  {journal} {\bibinfo
  {journal} {Journal of Superconductivity and Novel Magnetism}\ }\textbf
  {\bibinfo {volume} {30}},\ \bibinfo {pages} {1263} (\bibinfo {year}
  {2017})}\BibitemShut {NoStop}%
\bibitem [{\citenamefont {Lai}\ \emph {et~al.}(2008)\citenamefont {Lai},
  \citenamefont {Ho}, \citenamefont {Chang},\ and\ \citenamefont
  {Lin}}]{LaiHo(08)}%
  \BibitemOpen
  \bibfield  {author} {\bibinfo {author} {\bibfnamefont {Y.~H.}\ \bibnamefont
  {Lai}}, \bibinfo {author} {\bibfnamefont {J.~H.}\ \bibnamefont {Ho}},
  \bibinfo {author} {\bibfnamefont {C.~P.}\ \bibnamefont {Chang}}, \ and\
  \bibinfo {author} {\bibfnamefont {M.~F.}\ \bibnamefont {Lin}},\ }\href
  {\doibase 10.1103/PhysRevB.77.085426} {\bibfield  {journal} {\bibinfo
  {journal} {Phys. Rev. B}\ }\textbf {\bibinfo {volume} {77}},\ \bibinfo
  {pages} {085426} (\bibinfo {year} {2008})}\BibitemShut {NoStop}%
\bibitem [{\citenamefont {Bittencourt}\ and\ \citenamefont
  {Bernardini}(2017)}]{BitVic(17)}%
  \BibitemOpen
  \bibfield  {author} {\bibinfo {author} {\bibfnamefont {V.~A. S.~V.}\
  \bibnamefont {Bittencourt}}\ and\ \bibinfo {author} {\bibfnamefont {A.~E.}\
  \bibnamefont {Bernardini}},\ }\href {\doibase 10.1103/PhysRevB.95.195145}
  {\bibfield  {journal} {\bibinfo  {journal} {Phys. Rev. B}\ }\textbf {\bibinfo
  {volume} {95}},\ \bibinfo {pages} {195145} (\bibinfo {year}
  {2017})}\BibitemShut {NoStop}%
\bibitem [{\citenamefont {Ashhab}\ \emph {et~al.}(2007)\citenamefont {Ashhab},
  \citenamefont {Johansson}, \citenamefont {Zagoskin},\ and\ \citenamefont
  {Nori}}]{AshJoh(07)}%
  \BibitemOpen
  \bibfield  {author} {\bibinfo {author} {\bibfnamefont {S.}~\bibnamefont
  {Ashhab}}, \bibinfo {author} {\bibfnamefont {J.~R.}\ \bibnamefont
  {Johansson}}, \bibinfo {author} {\bibfnamefont {A.~M.}\ \bibnamefont
  {Zagoskin}}, \ and\ \bibinfo {author} {\bibfnamefont {F.}~\bibnamefont
  {Nori}},\ }\href {\doibase 10.1103/PhysRevA.75.063414} {\bibfield  {journal}
  {\bibinfo  {journal} {Phys. Rev. A}\ }\textbf {\bibinfo {volume} {75}},\
  \bibinfo {pages} {063414} (\bibinfo {year} {2007})}\BibitemShut {NoStop}%
\bibitem [{\citenamefont {Sen}\ \emph {et~al.}(2021)\citenamefont {Sen},
  \citenamefont {Sen},\ and\ \citenamefont {Sengupta}}]{SenSen(21)}%
  \BibitemOpen
  \bibfield  {author} {\bibinfo {author} {\bibfnamefont {A.}~\bibnamefont
  {Sen}}, \bibinfo {author} {\bibfnamefont {D.}~\bibnamefont {Sen}}, \ and\
  \bibinfo {author} {\bibfnamefont {K.}~\bibnamefont {Sengupta}},\ }\href
  {\doibase 10.1088/1361-648x/ac1b61} {\bibfield  {journal} {\bibinfo
  {journal} {Journal of Physics: Condensed Matter}\ }\textbf {\bibinfo {volume}
  {33}},\ \bibinfo {pages} {443003} (\bibinfo {year} {2021})}\BibitemShut
  {NoStop}%
\bibitem [{\citenamefont {Eckardt}(2017)}]{Eck(17)}%
  \BibitemOpen
  \bibfield  {author} {\bibinfo {author} {\bibfnamefont {A.}~\bibnamefont
  {Eckardt}},\ }\href {\doibase 10.1103/RevModPhys.89.011004} {\bibfield
  {journal} {\bibinfo  {journal} {Rev. Mod. Phys.}\ }\textbf {\bibinfo {volume}
  {89}},\ \bibinfo {pages} {011004} (\bibinfo {year} {2017})}\BibitemShut
  {NoStop}%
\bibitem [{\citenamefont {Zenesini}\ \emph {et~al.}(2009)\citenamefont
  {Zenesini}, \citenamefont {Lignier}, \citenamefont {Ciampini}, \citenamefont
  {Morsch},\ and\ \citenamefont {Arimondo}}]{ZenLig(09)}%
  \BibitemOpen
  \bibfield  {author} {\bibinfo {author} {\bibfnamefont {A.}~\bibnamefont
  {Zenesini}}, \bibinfo {author} {\bibfnamefont {H.}~\bibnamefont {Lignier}},
  \bibinfo {author} {\bibfnamefont {D.}~\bibnamefont {Ciampini}}, \bibinfo
  {author} {\bibfnamefont {O.}~\bibnamefont {Morsch}}, \ and\ \bibinfo {author}
  {\bibfnamefont {E.}~\bibnamefont {Arimondo}},\ }\href {\doibase
  10.1103/PhysRevLett.102.100403} {\bibfield  {journal} {\bibinfo  {journal}
  {Phys. Rev. Lett.}\ }\textbf {\bibinfo {volume} {102}},\ \bibinfo {pages}
  {100403} (\bibinfo {year} {2009})}\BibitemShut {NoStop}%
\bibitem [{\citenamefont {Ivanov}\ \emph {et~al.}(2008)\citenamefont {Ivanov},
  \citenamefont {Alberti}, \citenamefont {Schioppo}, \citenamefont {Ferrari},
  \citenamefont {Artoni}, \citenamefont {Chiofalo},\ and\ \citenamefont
  {Tino}}]{IvaSch(08)}%
  \BibitemOpen
  \bibfield  {author} {\bibinfo {author} {\bibfnamefont {V.~V.}\ \bibnamefont
  {Ivanov}}, \bibinfo {author} {\bibfnamefont {A.}~\bibnamefont {Alberti}},
  \bibinfo {author} {\bibfnamefont {M.}~\bibnamefont {Schioppo}}, \bibinfo
  {author} {\bibfnamefont {G.}~\bibnamefont {Ferrari}}, \bibinfo {author}
  {\bibfnamefont {M.}~\bibnamefont {Artoni}}, \bibinfo {author} {\bibfnamefont
  {M.~L.}\ \bibnamefont {Chiofalo}}, \ and\ \bibinfo {author} {\bibfnamefont
  {G.~M.}\ \bibnamefont {Tino}},\ }\href {\doibase
  10.1103/PhysRevLett.100.043602} {\bibfield  {journal} {\bibinfo  {journal}
  {Phys. Rev. Lett.}\ }\textbf {\bibinfo {volume} {100}},\ \bibinfo {pages}
  {043602} (\bibinfo {year} {2008})}\BibitemShut {NoStop}%
\bibitem [{\citenamefont {Grimm}\ \emph {et~al.}(1999)\citenamefont {Grimm},
  \citenamefont {Weidemüller},\ and\ \citenamefont
  {Ovchinnikov}}]{GriWei(99)}%
  \BibitemOpen
  \bibfield  {author} {\bibinfo {author} {\bibfnamefont {R.}~\bibnamefont
  {Grimm}}, \bibinfo {author} {\bibfnamefont {M.}~\bibnamefont {Weidemüller}},
  \ and\ \bibinfo {author} {\bibfnamefont {Y.~B.}\ \bibnamefont
  {Ovchinnikov}},\ }\href@noop {} {\enquote {\bibinfo {title} {Optical dipole
  traps for neutral atoms},}\ } (\bibinfo {year} {1999}),\ \Eprint
  {http://arxiv.org/abs/physics/9902072} {arXiv:physics/9902072 [physics.
  atom-ph]} \BibitemShut {NoStop}%
\bibitem [{\citenamefont {Bloch}\ \emph {et~al.}(2008)\citenamefont {Bloch},
  \citenamefont {Dalibard},\ and\ \citenamefont {Zwerger}}]{BloDal(08)}%
  \BibitemOpen
  \bibfield  {author} {\bibinfo {author} {\bibfnamefont {I.}~\bibnamefont
  {Bloch}}, \bibinfo {author} {\bibfnamefont {J.}~\bibnamefont {Dalibard}}, \
  and\ \bibinfo {author} {\bibfnamefont {W.}~\bibnamefont {Zwerger}},\ }\href
  {\doibase 10.1103/revmodphys.80.885} {\bibfield  {journal} {\bibinfo
  {journal} {Reviews of Modern Physics}\ }\textbf {\bibinfo {volume} {80}},\
  \bibinfo {pages} {885} (\bibinfo {year} {2008})}\BibitemShut {NoStop}%
\bibitem [{\citenamefont {Zhu}\ \emph {et~al.}(2007)\citenamefont {Zhu},
  \citenamefont {Wang},\ and\ \citenamefont {Duan}}]{ShiWan(07)}%
  \BibitemOpen
  \bibfield  {author} {\bibinfo {author} {\bibfnamefont {S.-L.}\ \bibnamefont
  {Zhu}}, \bibinfo {author} {\bibfnamefont {B.}~\bibnamefont {Wang}}, \ and\
  \bibinfo {author} {\bibfnamefont {L.-M.}\ \bibnamefont {Duan}},\ }\href
  {\doibase 10.1103/PhysRevLett.98.260402} {\bibfield  {journal} {\bibinfo
  {journal} {Phys. Rev. Lett.}\ }\textbf {\bibinfo {volume} {98}},\ \bibinfo
  {pages} {260402} (\bibinfo {year} {2007})}\BibitemShut {NoStop}%
\bibitem [{\citenamefont {Shao}\ \emph {et~al.}(2008)\citenamefont {Shao},
  \citenamefont {Zhu}, \citenamefont {Sheng}, \citenamefont {Xing},\ and\
  \citenamefont {Wang}}]{ShaShi(08)}%
  \BibitemOpen
  \bibfield  {author} {\bibinfo {author} {\bibfnamefont {L.~B.}\ \bibnamefont
  {Shao}}, \bibinfo {author} {\bibfnamefont {S.-L.}\ \bibnamefont {Zhu}},
  \bibinfo {author} {\bibfnamefont {L.}~\bibnamefont {Sheng}}, \bibinfo
  {author} {\bibfnamefont {D.~Y.}\ \bibnamefont {Xing}}, \ and\ \bibinfo
  {author} {\bibfnamefont {Z.~D.}\ \bibnamefont {Wang}},\ }\href {\doibase
  10.1103/PhysRevLett.101.246810} {\bibfield  {journal} {\bibinfo  {journal}
  {Phys. Rev. Lett.}\ }\textbf {\bibinfo {volume} {101}},\ \bibinfo {pages}
  {246810} (\bibinfo {year} {2008})}\BibitemShut {NoStop}%
\bibitem [{\citenamefont {Wu}\ \emph {et~al.}(2013)\citenamefont {Wu},
  \citenamefont {He}, \citenamefont {Zang},\ and\ \citenamefont
  {Kou}}]{Wuhe(13)}%
  \BibitemOpen
  \bibfield  {author} {\bibinfo {author} {\bibfnamefont {Y.-J.}\ \bibnamefont
  {Wu}}, \bibinfo {author} {\bibfnamefont {J.}~\bibnamefont {He}}, \bibinfo
  {author} {\bibfnamefont {C.-L.}\ \bibnamefont {Zang}}, \ and\ \bibinfo
  {author} {\bibfnamefont {S.-P.}\ \bibnamefont {Kou}},\ }\href {\doibase
  10.1140/epjb/e2013-31045-1} {\bibfield  {journal} {\bibinfo  {journal} {The
  European Physical Journal B}\ }\textbf {\bibinfo {volume} {86}} (\bibinfo
  {year} {2013}),\ 10.1140/epjb/e2013-31045-1}\BibitemShut {NoStop}%
\bibitem [{\citenamefont {Iba\~nez Azpiroz}\ \emph {et~al.}(2013)\citenamefont
  {Iba\~nez Azpiroz}, \citenamefont {Eiguren}, \citenamefont {Bergara},
  \citenamefont {Pettini},\ and\ \citenamefont {Modugno}}]{JulEig(13)}%
  \BibitemOpen
  \bibfield  {author} {\bibinfo {author} {\bibfnamefont {J.}~\bibnamefont
  {Iba\~nez Azpiroz}}, \bibinfo {author} {\bibfnamefont {A.}~\bibnamefont
  {Eiguren}}, \bibinfo {author} {\bibfnamefont {A.}~\bibnamefont {Bergara}},
  \bibinfo {author} {\bibfnamefont {G.}~\bibnamefont {Pettini}}, \ and\
  \bibinfo {author} {\bibfnamefont {M.}~\bibnamefont {Modugno}},\ }\href
  {\doibase 10.1103/PhysRevA.87.011602} {\bibfield  {journal} {\bibinfo
  {journal} {Phys. Rev. A}\ }\textbf {\bibinfo {volume} {87}},\ \bibinfo
  {pages} {011602} (\bibinfo {year} {2013})}\BibitemShut {NoStop}%
\end{thebibliography}%

\end{document}